\documentclass[sigplan,screen]{acmart}
\usepackage[normalem]{ulem}
\usepackage{booktabs}
\usepackage{threeparttable}
\usepackage{multirow}
\usepackage{enumitem}
\usepackage{mathtools}
\usepackage[ruled,vlined,linesnumbered]{algorithm2e}

\newcommand{\yyy}[1]{\textcolor{black}{#1}}

\copyrightyear{2024}
\acmYear{2024}
\setcopyright{rightsretained}
\acmConference[ASPLOS '24]{29th ACM International Conference on Architectural Support for Programming Languages and Operating Systems, Volume 2}{April 27-May 1, 2024}{La Jolla, CA, USA}
\acmBooktitle{29th ACM International Conference on Architectural Support for Programming Languages and Operating Systems, Volume 2 (ASPLOS '24), April 27-May 1, 2024, La Jolla, CA, USA}
\acmDOI{10.1145/3620665.3640383}
\acmISBN{979-8-4007-0385-0/24/04}

\begin{document}

\title{ExeGPT: Constraint-Aware Resource Scheduling for LLM Inference}

\author{Hyungjun Oh$^1$\quad Kihong Kim$^1$\quad Jaemin Kim$^1$\quad Sungkyun Kim$^1$
\\Junyeol Lee$^1$\quad Du-seong Chang$^2$\quad Jiwon Seo$^1$}
\def \authors{Hyungjun Oh, Kihong Kim, Jaemin Kim, Sungkyun Kim, Junyeol Lee, Du-seong Chang, Jiwon Seo}
\authornote{Corresponding author and principal investigator}
\affiliation{%
 \institution{$^1$Hanyang University\quad $^2$KT Corporation}
 \country{}
}
\affiliation{%
  \institution{Corresponding author: seojiwon@hanyang.ac.kr}
  \country{}
}

\thispagestyle{empty}

\begin{abstract}
This paper presents ExeGPT, a distributed system designed for constraint-aware LLM inference. ExeGPT finds and runs with an optimal execution schedule to maximize inference throughput while satisfying a given latency constraint. By leveraging the distribution of input and output sequences, it effectively allocates resources and determines optimal execution configurations, including batch sizes and partial tensor parallelism. We also introduce two scheduling strategies based on Round-Robin Allocation and Workload-Aware Allocation policies, suitable for different NLP workloads.

We evaluate ExeGPT on six LLM instances of T5, OPT, and GPT-3 and five NLP tasks, each with four distinct latency constraints. Compared to FasterTransformer, ExeGPT achieves up to 15.2$\times$ improvements in throughput and 6$\times$ improvements in latency. Overall, ExeGPT achieves an average throughput gain of 2.9$\times$ across twenty evaluation scenarios.  Moreover, when adapting to changing sequence distributions, the cost of adjusting the schedule in ExeGPT is reasonably modest. 
ExeGPT proves to be an effective solution for optimizing and executing LLM inference for diverse NLP workload and serving conditions.

\end{abstract}



\maketitle
\pagestyle{plain}

\section{Introduction}

\begin{figure*}
    \includegraphics[width=\textwidth]{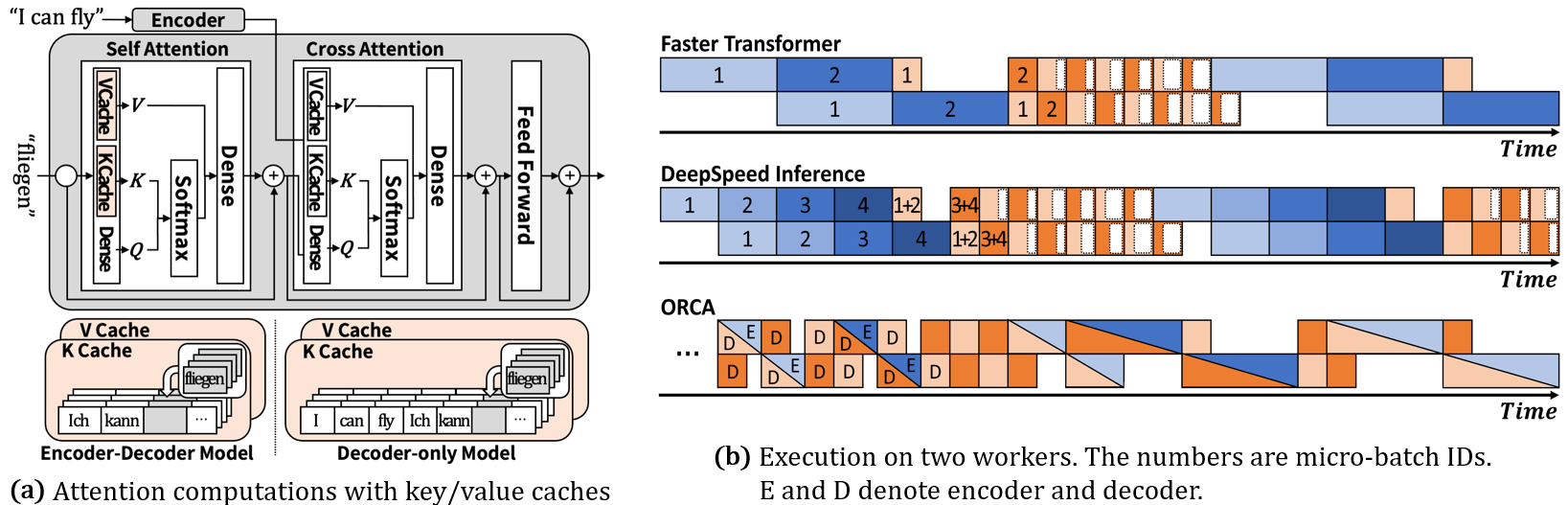}
    \caption{Illustration of a Transformer decoder (left) and inference timelines of three LLM inference systems (right).}
    \label{fig:background}
\end{figure*}

Large language models (LLMs) have significantly advanced the field of natural language processing (NLP) and enabled a wide range of NLP applications. However, their high computational costs make it challenging to run LLMs efficiently, limiting their full potential. For example, generating a single token in LLMs can require hundreds of billions of FLOPs, demonstrating the need for efficient execution.

Compared to other neural networks, LLM inference is challenging due to their large size and irregular executions. LLMs can have hundreds of billions of parameters, requiring model parallel execution on multiple GPUs. The autoregressive nature of LLM inference also complicates its execution, as each model execution generates a single token, which is fed back to generate the next one, requiring multiple iterations for completion. However, when executing for a batch of data, each input may need a different number of iterations, leading to much reduced batch sizes in later iterations. This workload variability makes it hard to maintain high throughput.

To mitigate this problem, ORCA proposed {\it iteration-level} scheduling that adds a new input data into a running batch to replace a completed one\,\cite{yu2022orca}. 
Although this ensures consistent batch sizes throughout execution, it does not consider that the cost of input encoding is orders of magnitude higher than that of output decoding, potentially causing large pipeline bubbles. Moreover, executing input encoding at any decoding iteration may result in highly variable and thus uncontrollable latency.

We suppose that NLP applications have diverse requirements for inference latency, with some needing immediate responses while others can tolerate longer delays. To achieve efficient inference serving under a given latency bound, it is essential to optimize resource utilization and computation throughput. However, the irregular workload between inputs makes it difficult to efficiently run LLM inference.

This paper introduces ExeGPT, a system designed for constraint-aware LLM inference. ExeGPT finds and runs with an optimal execution schedule that maximizes the inference throughput while satisfying a given latency constraint. It leverages the distribution of sequence lengths to efficiently allocate resources between GPUs and optimize the parallel configuration. The system provides two novel scheduling strategies with four control variables, such as decoding batches or partial tensor parallelism. The variables are designed to have a monotonic property, which we exploit in our scheduling algorithm to find an optimal execution schedule. This schedule is enforced by our distributed runner based on FasterTransformer.

We evaluated ExeGPT and other cutting-edge inference systems with six LLM configurations and five NLP tasks. 
ExeGPT shows significant improvements in throughput, up to 15.2$\times$, and latency, up to 6$\times$, when compared to FasterTransformer. Our contributions are summarized as follows.

\vspace{1pt}
\noindent
{\bf Scheduling Strategies and Latency/Throughput Trade-off.} 
We propose novel scheduling strategies based on two resource allocation policies, i.e., Round-Robin Allocation (RRA) and Workload-Aware Allocation (WAA). These strategies decouple the execution of encoding and decoding, allowing for efficient optimization of each phase. The strategies offer four control variables,  
enabling flexible trade-offs between throughput and latency for different workloads and applications. The scheduling strategies and the trade-off mechanism form the basis of our optimal scheduling algorithm.

\vspace{1pt}
\noindent
{\bf Scheduling with Sequence Length Distribution.} Our scheduler incorporates input and output sequence distributions, which can be obtained by observing NLP services over time, to determine the optimal schedule for real-world NLP workloads. We conducted probabilistic analysis to estimate the number of completed queries in decoding iterations and also their execution times using the profiling results. With the estimated execution times, we accurately simulate the timeline for the two strategies and find the optimal execution schedule.

\vspace{1pt}
\noindent
{\bf Optimizing Scheduling Algorithm.} 
We formulated an optimization problem for LLM inference to maximize throughput under a given latency bound by finding the optimal values of the control variables for RRA and WAA Scheduling. The optimization problem is monotonic, as the control variables are designed to be monotonic with respect to the throughput and latency. This allows us to implement an efficient optimization algorithm based on the standard branch-and-bound method. Our scheduling algorithm shows significant improvements in throughput and latency over existing methods.

\noindent
{\bf Extensive Evaluation.} 
We evaluated ExeGPT with six LLM instances of T5, OPT, and GPT-3 and five NLP tasks of summarization, translation, code generation, and conversational Q/A. For each task, we evaluated with four distinct latency constraints and measured the throughput. Compared to FasterTransformer, the state-of-the-art inference system, ExeGPT achieves up to 15.2$\times$ improvement in throughput and 6$\times$ in latency. On average, ExeGPT achieves 2.9$\times$ higher throughput across all the LLMs and tasks. Also, we observed that when adapting to changing sequence distributions, the cost of adjusting the schedule in ExeGPT is reasonably modest.

The paper is organized as follows. Section 2 provides background on LLM inference. Section 3 presents an overview of ExeGPT and its key components. Section 4 introduces our proposed scheduling strategies and their latency/throughput trade-off mechanism. In Section 5, we formulate the optimization problem for LLM inference and propose an efficient scheduling algorithm.  
Section 6 describes our simulation of execution schedules using the sequence length distributions. Section 7 evaluates ExeGPT and compares it with state-of-the-art systems.  
Finally, Section 8 concludes.

\section{Background and Related Work}
\label{sec:background}

LLMs are increasingly used in various NLP tasks, such as language translation, text summarization, and more. For all these tasks, LLM inference works by encoding the input once and then decoding the output in multiple autoregressive iterations,

with each decoding iteration producing a single output token. In Transformer-based models, the encoding and decoding phases are executed by corresponding layers\,\cite{vaswani2017attention}, while decoder-only models use decoding layers for both phases\,\cite{radford2018improving}. 

In these models, the attention mechanism captures token relationships in input and output sequences. It transforms each token into query, key, and value representations, computes the dot product and softmax values of the query and key vectors. The resulting attention weights are used to compute a context vector for each token, which is passed through a series of transformations, including a feedforward network and residual connections, to produce the layer output. The output is passed to the next layer, where the same computations are repeated.

In decoding computation of Transformer-based models, attention is computed in two ways: self-attention and cross-attention. Self-attention calculates attention scores between generated output tokens, while cross-attention calculates scores between each output token and all input tokens. \yyy{Figure\,\ref{fig:background}(a) shows 
a simplified overview of these computations in a decoding layer.} In decoder-only models, cross-attention layers are absent, and the decoders instead compute self-attention between all tokens in the input and output sequences.

During LLM inference, input and output sequences are used to generate the next token, which can result in redundant computations for previously generated tokens. To address this issue, fairseq\,\cite{ott2019fairseq} proposed incremental decoding techniques that memoize the keys and values of previous tokens. This memorization allows subsequent decoding iterations to only compute with a single token, thereby reducing the computation required for each decoding iteration. \yyy{Figure\,\ref{fig:background}(a) illustrates the memorization cache for self-attention at the bottom when decoding with token `fliegen', separately for Transformer-based models and decoder-only models.} Recently, vLLM\,\cite{vllm2023PagedAttention} proposed {\it PagedAttention} and applied paging to the key/value memorization cache for efficient memory management.

\yyy{We now review existing work on LLM inference optimization. While various approaches exist, including model compression techniques\,\cite{bondarenko2021understanding, dettmers2022llm, frantar2022gptq, tao2022compression, yao2022zeroquant, zafrir2019q8bert, zafrir2021prune, zhang2022platon, xiao2022smoothquant} and kernel acceleration techniques\,\cite{nvidia2021fastertransformer, mishra2021accelerating, dao2022flashattention}, we primarily focus on resource scheduling optimizations as they are closely related to our work. Following this review, we will discuss the problems associated with existing scheduling optimizations.}

\vspace{1pt}
\noindent
{\bf Review of Resource Scheduling Optimization.}
Due to the large number of parameters, LLMs require parallelism strategies to run their inference on multiple GPUs. One such approach is pipeline parallelism\,\cite{huang2019gpipe, li21chimera, li2021terapipe, narayanan19pipedream}, which splits an LLM by its encoding/decoding layers to create pipeline stages and employs micro-batching to reduce pipeline bubbles. Tensor parallelism\,\cite{shoeybi2019megatron} splits layers based on input/output tensors and corresponding parameters, for the parallel execution. This method requires synchronizing partial output tensors after each computation. Megatron\,\cite{narayanan2021efficient} has minimized this overhead and employs only two all-reduce synchronizations for an encoding layer and three for a decoding layer.

FasterTransformer\,(FT)\,\cite{nvidia2021fastertransformer} is NVIDIA's LLM inference engine. It allows to specify the degree of pipeline/tensor parallelism\,(PP and TP, respectively) and applies it to partition the encoders/decoders evenly across pipeline stages for the inference. Figure\,\ref{fig:background} shows FT's execution timeline for two GPUs with a (PP=2, TP=1) setting, running encoding once and decoding four iterations with two micro-batches. In the decoding iterations, FT maintains fixed batch sizes and does not early-terminate completed queries, resulting in unnecessary computation for those queries. This overhead is denoted by white boxes within decoding iterations. 

DeepSpeed Inference\,(DSI)\,\cite{aminabadi22dsi} is an LLM inference system. It applies low-level system optimizations, such as custom GeMM kernels for small batches, to optimize resource utilization. DSI also employs hybrid scheduling with different micro-batch sizes for encoding and decoding, using more micro-batches for encoding to reduce pipeline bubbles and fewer micro-batches for decoding to improve throughput. Figure\,\ref{fig:background} shows DSI's timeline in the middle running with four micro-batches for encoding and two for decoding. The technique of adjusting micro-batches in this way is now adopted in FT. DSI maintains fixed batch sizes during decoding, which leads to unnecessary computation as in FT. While it is possible to early-terminate completed queries, that results in reduced batch sizes, making it difficult to maintain high throughput.

ORCA\,\cite{yu2022orca} is an LLM inference engine and serving system. It proposes iteration-level scheduling, that combines ongoing and new requests in the same batch and executes their encoding and decoding iterations simultaneously. This approach requires decoding for a batch of data having different number of generated tokens. Thus, ORCA executes attention computations at the granularity of individual requests and performs the rest of encoding/decoding computations at the batch granularity. With iteration-level scheduling, ORCA can apply early termination of completed requests while still maintaining the same batch sizes and computation throughput. Figure\,\ref{fig:background} shows the timeline of ORCA running encoding and decoding in the same micro-batches. Their scheduling, however, does not consider the difference in the amount of computation between encoding and decoding, which may result in large pipeline bubbles as shown in the figure.

\vspace{1pt}
\noindent
{\bf \textcolor{black}{Limitations} in Current Scheduling Optimization Work.}
\textcolor{black}{Existing work suffers from} the performance degradation problem due to decreasing decoding batches (in FT and DSI) or pipeline bubbles (in ORCA). 
In the former case, queries in a same batch may require varying numbers of decoding iterations for completion. Hence, as decoding proceeds, batch sizes decrease, leading to a degradation of hardware resource utilization and the inference throughput.

To maintain large decoding batches, ORCA replaces completed queries with encoding of new queries within the same batch as the decoding of previous queries. However, due to differences in encoding and decoding workloads, this may induce pipeline bubbles, as shown in Figure\,\ref{fig:background}(b).

Another issue in the existing inference systems is their inefficient trade-off mechanism between throughput and latency. To reduce the inference latency, these systems resort to reducing batch sizes, which can significantly undermine the throughput in certain parallel settings. Moreover, in the case of ORCA, the encoding of new queries may occur in any decoding iteration, affecting the latency of ongoing queries and making it challenging to control inference latency.

\section{Overview of ExeGPT}
\label{sec:overview}

\begin{figure}
\centering
    \includegraphics[width=0.48\textwidth]{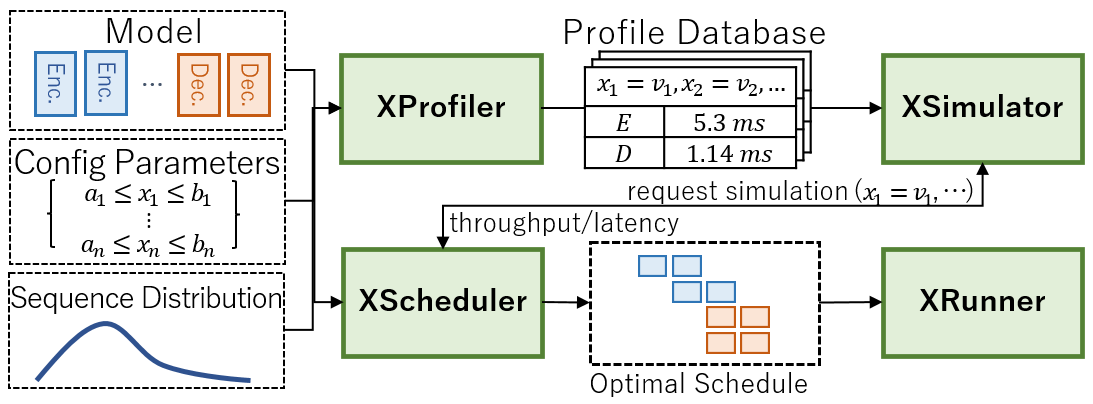}
    \caption{High-level architecture of ExeGPT}
    \label{fig:overview}
\end{figure}

ExeGPT is a system for running LLM inference with optimal execution schedule. 
Under a given latency constraint, it finds and runs with the execution schedule that maximizes inference throughput. ExeGPT offers efficient methods for throughput and latency trade-off ensuring that the execution maintains adequate decoding batches without introducing substantial pipeline bubbles.

The system comprises four primary components: XProfiler, XSimulator, XScheduler, and XRunner, which are illustrated in Figure\,\ref{fig:overview}. 
XProfiler measures the execution times of a single encoding and decoding layer across all possible configurations. With the execution times of a single layer, XSimulator constructs the timeline when requested by XScheduler. 
The scheduler identifies optimal configuration parameter values using the simulator to estimate throughput and latency at selected configuration points.
XRunner then executes with the configuration parameters determined by the scheduler.
Below we provide further details of each of these four system components.

\vspace{1pt}
\noindent
{\bf XProfiler.} 
For a single encoding and decoding layer, the profiler separately measures the execution times of the attention kernel and the rest of the encoding/decoding layer, considering all feasible tensor-parallel degrees. 
For the former (the attention kernel), we conduct sweeps across batch sizes and, for each batch size, perform sweeps over possible sequence lengths.
For the latter, we measure the execution time with sweeping input sizes (i.e., batch sizes$\times$input lengths). 

We also measure the synchronization overhead of tensor- and pipeline-parallel execution. These synchronization operations do not interfere with each other as they either occur at different time points (between tensor- and pipeline-parallel synchronization) or utilize point-to-point communication channels (between different tensor-parallel groups).

\vspace{1pt}
\noindent
{\bf XSimulator.} Using the profiling results, the simulator constructs the execution timeline based on the configuration parameters, such as batch size and tensor-parallel degree, given by the scheduler.
For the simulation, we calculate the expected batch sizes of each encoding/decoding iteration, utilizing the probability distribution of the input/output sequence lengths. 
Our scheduling algorithm allows queries from different encoding iterations to be decoded in a same batch. The simulator takes this into account and calculates the corresponding distribution for the timeline simulation (details explained in Section\,\ref{sec:simulator}). 

\vspace{1pt}
\noindent
{\bf XScheduler.} The scheduler searches over the solution space to find the optimal values of the configuration parameters that maximize throughput under a given latency bound. 
The details are explained in Section\,\ref{sec:optimizer},
but we have developed a branch-and-bound search algorithm 
specifically designed for the efficient exploration.
This algorithm exploits the monotonic property of configuration parameters with respect to throughput and latency, enabling to examine 
a small subset of the configuration points to pinpoint the optimal solution.

\vspace{1pt}
\noindent
{\bf XRunner.} This serves as the execution engine for running LLM inference on multiple distributed GPUs. It takes as input the execution schedule provided by XScheduler and ensures that the schedule is enforced during the execution.

To implement XRunner, we extended NVIDIA's FasterTransformer. We first implemented early-termination of completed queries in a batch, along with the compaction of the key/value cache entries for those terminated queries. For WAA Scheduling (explained in Section\,\ref{sec:strategy}), we implemented the transfer of the key/value cache entries from encoding GPUs to decoding GPUs. To minimize interference, we copy the entries from the source GPU memory to CPU memory and then transfer them to the target GPU memory.
We also implemented several optimizations, such as overlapping of computation and communication for pipeline-parallel execution.

\vspace*{-0.5\baselineskip}
\section{Scheduling Strategies and Trade-offs} \label{sec:strategy}

For efficient LLM inference, we must strike a balance between encoding and decoding operations. This balance must ensure maintaining sufficiently large decoding batches while also minimizing pipeline bubbles.
However, in current inference systems, these two operations are tightly coupled, hindering efficient execution. For instance, in all existing systems, there is an implicit assumption that the decoding of a query takes place immediately after its encoding, and it is further assumed that all decoding iterations of a query are executed consecutively. From a resource allocation perspective, GPUs that execute the $i$'th encoding layer are expected to handle the $i$'th decoding layer in all existing systems.

\begin{figure}[t!]
\centering
    \includegraphics[width=0.48\textwidth]{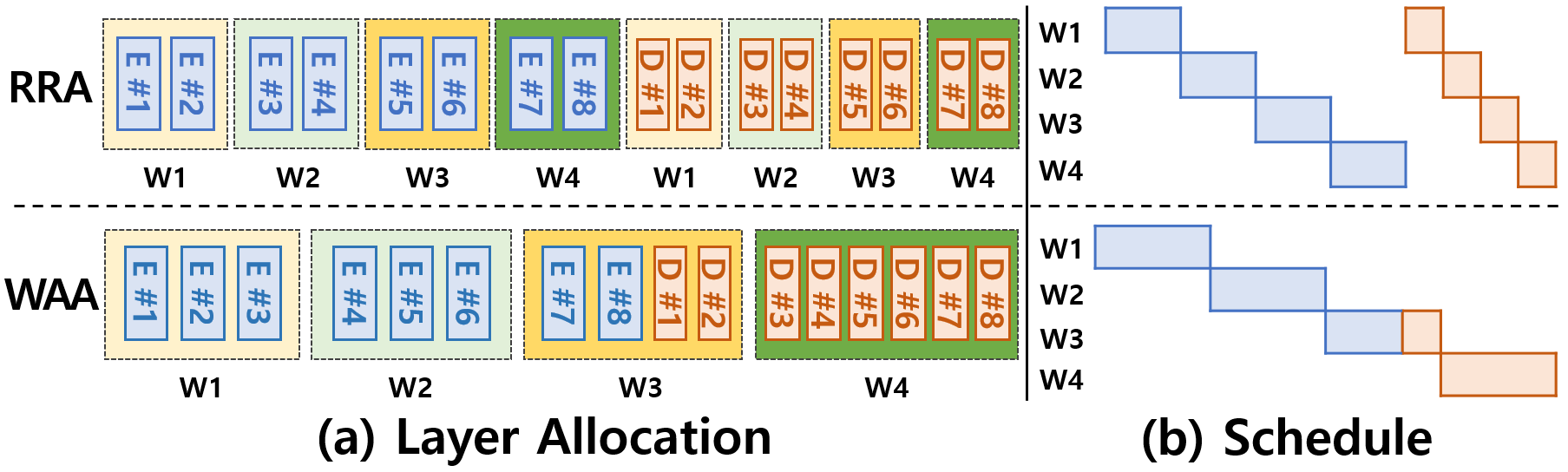}
    \caption{Comparing RRA and WAA Scheduling strategies} 
    \label{fig:allocation}
\end{figure}

In this section, we challenge these assumptions and propose two alternative scheduling strategies that decouple the execution of encoding and decoding differently, allowing for flexible resource allocation and efficient execution. Furthermore, we introduce four control mechanisms that offer a flexible trade-off between latency and throughput. These methods are employed by XScheduler's optimal scheduling algorithm, which is described in detail in Section\,\ref{sec:optimizer}.

\subsection{Decoupled Scheduling of Encoders and Decoders}
\label{sec:decoupled-scheduling}
We first present our layer allocation policies and then the scheduling strategies for each allocation. Our scheduling is based on two allocation policies: Round-Robin Allocation (RRA) and Workload-Aware Allocation (WAA). RRA simply assigns GPUs to the encoders and decoders of the model in a round-robin fashion. That is, when allocating for $N$ GPUs, RRA assigns $\frac{E}{N}$ consecutive encoders and $\frac{D}{N}$ consecutive decoders to each GPU, where $E$ and $D$ are the number of encoders and decoders respectively. Figure\,\ref{fig:allocation} illustrates this allocation at the top for 4 GPUs and 8 encoders/decoders.

The WAA policy estimates the workload sizes of the encoding/decoding layers and assigns GPUs proportionally to these sizes, with some GPUs dedicated to encoding and others to decoding in a pipelined manner.
It has two sub-types, WAA-C and WAA-M, which assign the layers based on their computation time and memory consumption, respectively.
When allocating for $N$ GPUs and $N$ pipeline stages, WAA-C assigns $[N\frac{C_E}{(C_E+C_D)}]$ GPUs to encoders and $[N\frac{C_D}{(C_E+C_D)}]$ GPUs to decoders, where $C_E$ and $C_D$ are estimated encoding/decoding times. This balances the amount of computation between encoding and decoding GPUs. Notably, we decouple the encoding and decoding stages and run them asynchronously to prevent delays caused by varying input lengths during encoding.
Figure\,\ref{fig:allocation} (bottom) shows an example of WAA policy.

WAA-M assigns GPUs to ensure that all GPUs consume an equal amount of memory rather than equal computation. In WAA-C the decoding GPUs consume more memory due to larger key/value caches for storing output tokens. When this becomes the performance bottleneck, WAA-M can be used to improve the performance by, for example, increasing the batch size. However, using WAA-M may result in much longer encoding stages, as fewer GPUs are assigned to encoding. This increases the worst-case latency, which we consider in our scheduling to meet latency constraints.

For a decoder-only model, where a decoding layer runs both input encoding and output decoding, WAA stores two copies of the model as it dedicates each GPU for either encoding or decoding. 
This memory overhead is reasonably modest compared to the key/value cache sizes, and we provide a detailed evaluation in Section\,\ref{sec:eval-small-llm}. We now describe the scheduling strategies. 

\begin{figure}
\centering
    \includegraphics[width=0.48\textwidth]{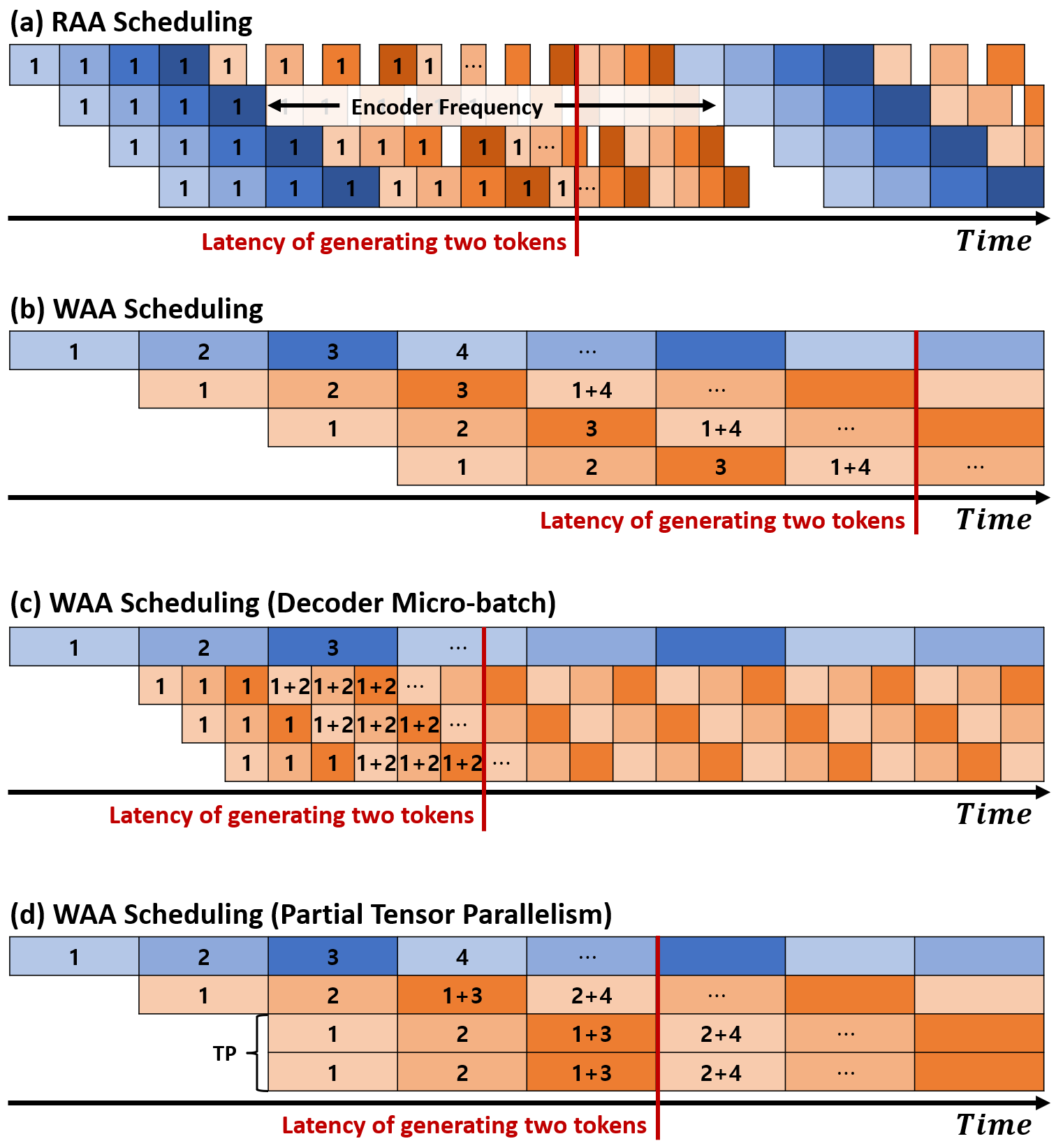}
    \caption{Execution timelines of RRA/WAA Scheduling with GPU$_1$--GPU$_4$. The blue/red boxes are encoding/decoding iterations, respectively. The numbers are mini-batch IDs. 
    \label{fig:control_vars}}
\end{figure}

\noindent
{\bf RRA Scheduling.} This strategy executes input encoding once and then decoding for $N_D$ iterations, repeatedly executing the encoding and decoding phases as in Figure\,\ref{fig:control_vars}(a). By running the decoding only for a fixed number of iterations and then feeding in more input data, we keep the decoding batch size sufficiently large for high resource utilization. It is worth noting that running encoding more frequently (decreasing $N_D$) increases throughput but comes at the cost of longer latency. 

To configure RRA Scheduling, we need to set the encoding batch size $B_E$, 
decoding batch size $B_D$, and decoding iteration $N_D$. We set $B_E$ to match the number of completed queries in the previous $N_D$ decoding iterations to maintain consistent batch sizes across repeated encoding/decoding phases.

\noindent
{\bf WAA Scheduling.} This strategy executes encoders and decoders in their assigned GPUs, decoupling the two phases, as shown in Figure\,\ref{fig:control_vars}(b). The decoupling enables encoding and decoding to operate with different batch sizes. When an input batch is encoded and passed for decoding, it is merged and decoded with previously decoded batch data.

To configure WAA Scheduling, we need to determine the encoding and decoding batch sizes. To maintain consistent batch sizes, we use the output sequence length to set the batch sizes. That is, if the average length of output sequences is $S_D$, we set the decoding batch size $B_D$ equal to the encoding batch size $B_E$ multiplied by $S_D$ (i.e., $B_D\!=\!B_E\cdot S_D$). These batch sizes impact the encoding and decoding times of a pipeline stage, which is used for the allocation of layers by the WAA policy.

\noindent
{\bf Comparison of the Strategies.} 
RRA and WAA Scheduling both address the performance challenges in existing systems by simultaneously tackling the problems of diminishing decoding batches (in FT and DSI) and pipeline bubbles (in ORCA). 
RRA Scheduling places more emphasis on maintaining large decoding batches than minimizing pipeline bubbles, allowing for a small amount of bubbles between encoding and decoding phases as shown in Figure\,\ref{fig:control_vars}(a). 
In contrast, WAA Scheduling results in smaller encoding/decoding batches than those of RRA for decoder-only models due to the memory overhead. However, it excels in minimizing pipeline bubbles by separately optimizing the encoding and decoding stages.

Due to these differences, WAA performs better when the output sequence length is relatively short. In such cases, RRA requires more frequent execution of encoding phases, exacerbating the pipeline bubble problem.
Conversely, for large decoder-only models, RRA outperforms WAA because it does not incur the memory overhead associated with WAA. We discuss these differences further in our evaluation.

\subsection{Controlling Latency and Throughput Trade-off}

RRA and WAA provide four control mechanisms to tailor execution schedules: batch size, decoder micro-batch, partial tensor parallelism, and encoding frequency.
These variables enable to increase throughput at the cost of longer latency. 
Notably, three of the mechanisms, except batch size, are novel proposals in this paper. We explain these four mechanisms.

\noindent
{\bf Batch size.} Increasing the batch size improves the parallelism of the inference kernels and thus improves the throughput but it also increases the latency.
Hence the batch size is generally used for the trade off of inference throughput and latency.

\noindent
{\bf Decoder micro-batch.} In WAA, we make it possible to split a decoding batch into multiple micro-batches, which can largely reduce the latency.
Figure\,\ref{fig:control_vars}(b) and (c) illustrate the impact of applying decoder micro-batches. In Figure\,\ref{fig:control_vars}(b), with GPU$_1$ encoding and GPU$_2$--GPU$_4$ decoding, GPU$_2$ has to wait for GPU$_4$ to complete the decoding iteration before it can start the next iteration, thus requiring seven pipeline stages to generate two tokens.
In contrast, applying micro-batches, as shown in Figure\,\ref{fig:control_vars}(c), enables overlapping of decoding between micro-batches, requiring only 3$\frac{2}{3}$ pipeline stages for generating two tokens.
Decoder micro-batches can be used for the trade off of throughput and latency
as increasing the number of micro-batches reduces the latency but decreases the throughput.

\noindent
{\bf Partial tensor parallelism.} By splitting a layer and executing it on multiple GPUs in a tensor parallel way, we can reduce the depth of the execution pipeline and decrease latency at the cost of lower throughput due to synchronization overhead. 
Our approach allows tensor parallelism to be applied to a subset of GPUs, as shown in Figure,\ref{fig:control_vars}(d), enabling a trade-off between latency and throughput by increasing or decreasing the subset. When the degree of tensor parallelism is fixed, increasing the number of GPUs in tensor-parallel execution reduces latency by decreasing the pipeline depth but increases throughput.

\noindent
{\bf Encoding frequency.} In RRA Scheduling, adjusting the encoding frequency can increase throughput or decrease latency. That is, running the encoding more frequently maintains larger decoding batches, increasing throughput at the cost of latency.

XScheduler optimizes throughput under a given latency bound by adjusting these four variables. Its scheduling algorithm takes advantage of their characteristics of throughput and latency trade-off, which we explain in the next section.

\section{Constraint-Aware Scheduling Algorithm}
\label{sec:optimizer}
We assume that NLP applications have diverse latency constraints for generating a sequence within guaranteed timeframes. To maximize the resource utilization under these constraints, we formulate an optimization problem in this section and present an scheduling algorithm to address the problem.

Our optimization problem is to maximize the inference throughput under a given latency constraint by finding the optimal values of the control variables for the given scheduling policy. More formally the problem is stated as below.

\begin{equation*}
\small 
\begin{aligned}
 & \underset{B_{\{E, D, m\}}, T_P, F_E, S}
             {\text{arg\,max}} Throughput(B_E, B_D, B_m, T_P, F_E, S, P_E, P_D)   \\
 & \text{s.t.} ~ Latency(B_E, B_D, B_m, T_P, F_E, S, P_E, P_D) < L_{Bound} \text{,\phantom{12}\small{where}}\\
\end{aligned}
\end{equation*}

\begin{itemize}[leftmargin=0.1cm]
\item $B_E$, $B_D$, and $B_m$ are the sizes of encoder/decoder batches and decoder micro-batches,
\item $T_P$ is the tensor-parallelism degree and applied GPU count, 
\item $F_E$ is the frequency of running encoders for RRA Scheduling,
\item $S$ is the given scheduling policy (RRA, WAA-C, or WAA-M),
\item $P_E$ and $P_D$ are the given distributions of input/output length,
\item $Throughput()$ and $Latency()$ are the functions that give the throughput and latency for the given environment, and
\item $L_{Bound}$ is the latency bound for the given sequence length.
\end{itemize}

The problem can be solved by applying black-box optimization techniques such as Bayesian optimization\,\cite{movckus1975bayesian}, which do not rely on any assumptions about the objective functions. Alternatively, we can take advantage of the monotonicity property of the control variables for faster optimization by customizing existing algorithmic frameworks for monotonic optimization. We explain this approach in the following section.

\subsection{Scheduling Algorithm for Monotonic Optimization}
The optimization problem we formulated is monotonic\,\cite{tuy2000monotonic}, meaning the objective and constraint functions are monotonic with respect to each control variable.
This is because of the previously explained nature of the control variables and we evaluate monotonicity of the functions and report the results. The details are explained later but in short the functions are monotonic within small tolerance values.
To leverage this monotonicity property, we designed an optimization algorithm based on the branch-and-bound method\,\cite{tuy2006discrete}. It is a standard method for monotonic optimization problems, and is known to converge faster than the polyblock algorithm\,\cite{tuy2000monotonic}, an alternative approach for the problem. 

Our optimization algorithm is presented in Algorithm~1. For simplicity, we explain the algorithm using two control variables, but it can be simply extended to more variables.  The algorithm takes as input the ranges of the variables, the tolerance values for throughput and latency, and the latency bound. The tolerance values are used to find the optimal solution even when the objective and constraint functions are not monotonic in some sub-ranges of the search space. Smaller tolerance values lead to faster search with more pruning, but may result in a sub-optimal solution.

\begin{figure}
\centering
    \includegraphics[width=0.48\textwidth]{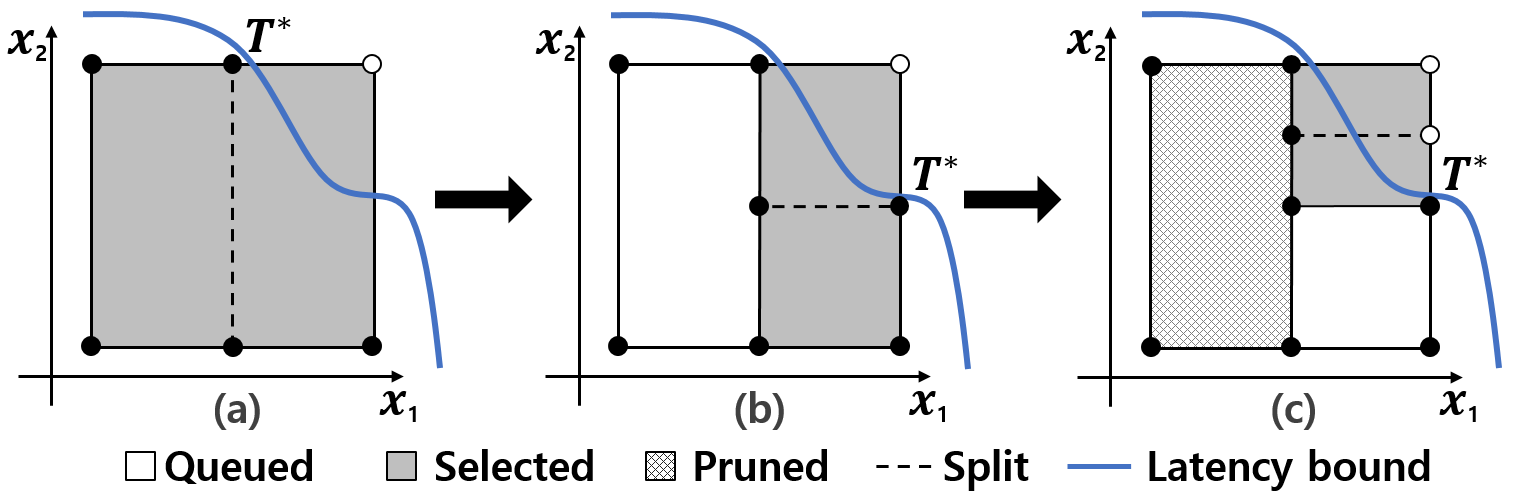}
    \caption{
    Illustration of the branch-and-bound scheduling. The filled circles are configuration points within the latency bound and empty circles are those that are not. $T^*$ is the current optimal point.
    }
    \label{fig:algorithm}
\end{figure}

The algorithm starts by considering the initial block $B_0$ given by the bounds on control variables $x_1$ and $x_2$. If the latency constraint is satisfied on the boundary when $x_1\!=\!b_1$ and $x_2\!=\!b_2$, then it is the optimal solution. Otherwise, the algorithm finds the optimal point within the block performing the following four steps: selecting a block, splitting the block, updating the current optimal point, and pruning the blocks in the queue. The steps are illustrated in Figure\,\ref{fig:algorithm} with an example. The throughput and latency of a block's boundary points are estimated using XSimulator (described in Section\,\ref{sec:simulator}) with the configuration of those points.

In the block splitting step, we use a heuristic method for faster convergence (lines 7--10). We examine the throughput and latency of the top left and bottom right points of the current block and identify the point with the higher throughput satisfying the latency constraint. If the top left point has a higher throughput, the block is split vertically; otherwise, it is split horizontally. If neither point satisfies the constraint, the block is split along its longer dimension.

These steps are repeated as long as there are blocks in the queue. The algorithm terminates when the queue is empty and the final solution to the problem is determined by the optimal throughput and its configuration at that time. We run the algorithm separately for RRA and WAA Schedule and select the solution that gives the highest throughput.

\SetKwInput{KwInput}{Input}
\SetKwInput{KwOutput}{Output}
\begin{algorithm}[t]
\small
\caption{Branch-and-bound Scheduling\label{algo:sched}}
\KwInput{$[a_1, b_1], [a_2, b_2]$: ranges of variables $x_1$, $x_2$,\\
\phantom{123456}$\epsilon_{T}, \epsilon_{L}$: tolerance for throughput and latency, \\
\phantom{123456}$L_{b}$: latency bound.
}
\KwOutput{Scheduling Configuration $x_1=v_1$, $x_2=v_2$} 
 $B_0 = [(a_1, a_2), (b_1, b_2)]$\;
 $B_0.lowr = perf(a_1, a_2)$; \,\,// gives (latency, throughput)\\
 $T^{*} = B_0.lowr.thrput$; \, $T^{*}.config=(a_1, a_2)$\;
 $Q = PriorityQueue(\{B\})$; // sorted by lower bound \\
 \While {$\mathcal{Q} \neq \emptyset$} {
   $B = Q.popMax()$\; 
   $p_{tl}, p_{br} = perf(B.topLeft), perf(B.bottomRight)$\;
   $perf_* \leftarrow$ higher thrput satisfying $L_{b}$ in $[p_{tl}, p_{br}]$\;
   \lIf {$perf_{*} == p_{tl}$} { $B_1, B_2 \leftarrow$ split B vertically}
   \lElse { $B_1, B_2 \leftarrow$ split B horizontally}
   \ForEach {$B' \in [B_1, B_2]$} {
      $B'.upp = perf(B'.topRight)$\;
      $B'.lowr = perf(B'.bottomLeft)$\;
      \lIf{ $B'.lowr.latency<L_{b}+\epsilon_{L}$} { $Q.add(B')$}
   }
   $B' \leftarrow$ higher throughput satisfying $L_{b}$ in $[B_1.upp, B_2.upp]$\;
   \If{$B'.upp.thrput\!>\!T^{*}$} { 
      $T^* = B'.upp.thrput$; \, $T^*.config = B'.topRight$\;
      $Q \leftarrow Q \setminus \{\hat{B} | \hat{B}.upp.thrput+\epsilon_{T} < T^{*} \}$
   }
 }
 \Return{$T\,^{*}.config$}\;
\end{algorithm}

When running the algorithm, we treat partial tensor parallelism specially. The variable controls two values: parallelism degree and the number of GPUs to which tensor parallelism is applied. To enforce the monotonicity property, we fix the parallelism degree as a constant and then run the algorithm. With a given degree, increasing the number of tensor-parallel GPUs reduces the latency and decreases the throughput, making the variable monotonic (which we verify in the evaluation section). We run the algorithm multiple times with different degree values to find the optimal solution.
 
\subsection{Dynamic Scheduling for Consistent Workload}

Both RRA and WAA Schedule keep average encoder/decoder batch sizes be consistent for their scheduling. However, the encoding and decoding workload can vary due to variations in input and output sequence lengths across different queries. In WAA, for example, a long encoding stage can miss the handover of its output to the decoding stage, potentially resulting in uneven decoding batches and execution times.

To achieve consistent and reliable inference execution, we implement dynamic workload adjustment at runtime. Specifically, we dynamically adjust the encoder batch size to ensure that the encoder workload (i.e., sum of input sequence lengths in a batch) stays within a predetermined threshold of the average workload. Moreover, to maintain consistent decoder workload, we monitor its batch size, and if it falls below/above the threshold compared to the average workload, we increase/decrease the encoder batch size accordingly. These adjustments are taken into account in our scheduling process.

\section{Simulation with Sequence Distribution}
\label{sec:simulator}

We implemented XSimulator, an execution simulator that utilizes the probability distribution of input and output sequence lengths. It incorporates profiling results to accurately estimate the execution times of encoding/decoding layers by evaluating a single encoder/decoder with all possible parallel configurations of the four variables, i.e., batch sizes, decoder micro-batches, partial tensor parallelism, and encoding frequency. 

As encoding and decoding are decoupled in our approach, queries from different encoding iterations may be decoded in a same batch. Hence, those queries in the batch are, on average, completed at different decoding iterations. We take this into account to maintain a consistent average encoding/decoding workload and achieve accurate scheduling results.

For both RRA and WAA Schedule, we assume probability distributions $P_E(S)$ and $P_D(S)$ for input and output sequence lengths, respectively. $P_E(S)$ and $P_D(S)$ may or may not be correlated, which we discuss with public datasets in our evaluation, and we assume they are uncorrelated in our simulator.
When input and output length are correlated, we can introduce randomization in the input lengths across different batches to mitigate potential biases by varying input lengths.

\noindent
{\bf Simulating RRA Schedule.} The simulator takes as input the encoder batch size $B_E$, the encoding frequency (i.e., the number of decoding iterations $N_D$ between encoding), and the sequence distributions $P_E(S)$ and $P_D(S)$. Using $P_D(S)$, we compute the distribution $P_D(U)$, the probability of completing the decoding of a query at $U$'th iteration after the previous encoding phase ($U$$\leq$$N_D$, details described below). It then sets the encoding batch size $B_E$ to $B_D \sum_{U} P_D(U)$, which is the expected number of completed queries after $N_D$ decoding iterations. This ensures that the encoding and decoding batch sizes are consistent across executions. Using the calculated batch sizes, the simulator estimates the expected encoding and decoding times of each pipeline stage accurately.

To compute $P_D(U)$, we first consider the conditional distribution $P_D(U|S\!=\!s)$, which represents the probability of completing the decoding of a query in $U$ iterations after the most recent encoding phase, given that the query generates a sequence of length $S$\,=\,$s$. $P_D(U|S)$ is calculated as follows.

{\small 
\[
P_D(U|S)\!=
\begin{cases}
    \begin{rcases}
        1, & \text{if } U=S  \\
        0, & \text{otherwise }  
    \end{rcases}, & \!\text{if } S\!\leq\!N_D \\
    \begin{rcases}
        \frac{1}{\lceil \frac{S}{N_D} \rceil}, & \text{if } U\!=\!1\!+\!(S\!-\!1\bmod N_D) \\
        0, &\!\text{otherwise} 
    \end{rcases}, & \text{if } S\!>\!N_D
\end{cases}
\]
}

\noindent
In the above, decoding a sequence of length $S$$>$$N_D$ requires $\lceil \frac{S}{N_D} \rceil$ decoding phases. 
Thus, the probability of completing the decoding in $U$ iterations during any specific decoding phase is calculated as $\frac{1}{\lceil \frac{S}{N_D} \rceil}$, with the complement representing the probability of not completing the decoding during that phase.

$P_D(U)$ is calculated as follows: $P_D(U)\!=\!\sum_{S} P_D(U|S)P_D(S)$.  Using $P_D(U)$, we set the decoding batch size $B_D$ to $\frac{B_E}{\sum_{U} P_D(U)}$. We then use $B_E$ and $P_E(S)$ to estimate the expected encoding workload, and $B_D$ and $P_D(U)$ to compute the expected decoding workload and their corresponding execution times. With these calculated times, the simulator creates an execution timeline and estimate the throughput and latency.

\noindent
{\bf Simulating WAA Schedule:} This is similar to the RRA case with encoder running once after each decoding iteration. As $N_D\!=\!1$, the probability of query completion at any decoding iteration in WAA is equivalent. Thus, we can simply use the average input and output sequence length ($S_E$ and $S_D$, respectively) to compute the workload sizes. We set the decoding batch size $B_D$ to be $B_E\cdot S_D$ for given $B_E$ to ensure consistent batch sizes across executions. Then the average workload sizes ($B_E\cdot S_E$ for encoding and $B_D$ for decoding) are used to estimate the total encoding and decoding times with the profile results. We allocate GPUs proportional to those times, as explained in Section\,\ref{sec:strategy}. 
The computed execution times, including buffer time for dynamic adjustments, are used to simulate the timeline and estimate throughput and latency.

\section{Evaluation}
\label{sec:eval}

In this section, we present the evaluation of our prototype ExeGPT system.
To evaluate the effectiveness of our scheduling strategies and constraint-aware scheduling algorithm, we conducted a comprehensive evaluation using six different LLM configurations across twenty scenarios. Moreover, we compared the evaluation results with those of FasterTransformer, DeepSpeed Inference (DSI), ORCA, and vLLM, which are state-of-the-art LLM inference systems. In the following sections, we describe our evaluation methodology and present the evaluation results.

\begin{table}[t]
\centering
\small
\caption{Evaluated Models and Configurations}
\label{tab:model}
\begin{tabular}{@{}lcccc@{}}
  \toprule
  \textbf{Model} & \textbf{\# Params} &  \textbf{\# Layers} & \textbf{Hidden Size} & \textbf{\# Atten. Head} \\
  \midrule
  \,T5 & 11B & 48 & 1024 & 128\\
  \cmidrule(l{1pt}r{\lightrulewidth}){1-5}
  \,OPT & 13B & 40 & 5120& 40\\
  \cmidrule(l{1pt}r{\lightrulewidth}){1-5}
  \multirow{4}{*}{\,GPT-3} & 39B & 48 & 8192& 64\\
                           & 101B & 80 & 10240& 80\\
                           & 175B & 96 & 12288& 96\\
                           & 341B & 120 & 15360& 120\\
  \bottomrule
\end{tabular}
\end{table}

\subsection{Evaluation Methodology}

{\bf Evaluated Models and System Settings.}  
The existing LLMs are primarily based on Transformer structures. Some models, such as T5\,\cite{raffel2020exploring} and UL2\,\cite{tay2022unifying}, consist of both encoders and decoders, while others, such as GPT-3\,\cite{brown2020language} and LaMDA\,\cite{thoppilan2022lamda}, are decoder-only models. To evaluate how our proposed techniques affect performance on different LLM configurations, 
we conducted experiments using representative models: T5, OPT\,\cite{zhang2022opt}, and GPT-3, covering both encoder-decoder and decoder-only models. We evaluated small to large versions of these models in half-precision (FP16) with various configurations as shown in Table\,\ref{tab:model}, which covers most model configurations in the ORCA and DSI papers\,\cite{aminabadi22dsi, yu2022orca}. Note that recent models like Gopher\,\cite{rae2021scaling}, LLaMA\,\cite{touvron2023llama, touvron2023llama2} and Alpaca\,\cite{taori2023alpaca} are either structurally equivalent to these models or very similar, with any minor differences resulting in the same amount of computation\,\cite{shazeer2020glu}.

We executed the inference of these models on two GPU clusters, the A40 cluster and A100 cluster as shown in Table\,\ref{tab:gpu}. The A40 cluster is a private cluster having six machines, each with eight A40 GPUs with 48GB of memory, for a total of 48 GPUs, connected via PCIe 4.0$\scriptstyle \times$16. The machines are connected via 100Gb Infiniband network. The A100 cluster consists of two NDm A100 v4 VMs on the Microsoft Azure cloud, each with eight A100 GPUs with 80GB of memory, for a total of 16 GPUs, connected via NVLink 3.0. The VMs are connected via Infiniband with 1.6Tb bandwidth between VMs using 8$\scriptstyle \times$200Gbp Mellanox HDR Infiniband adapters. 

Table\,\ref{tab:gpu} also shows the LLM configurations and the sub-clusters on which the models are executed. We tested the GPT-3 175B model on both the A40 and A100 clusters, as it is widely studied in NLP and other domains. While a subset of the models in the table can manage running with half the number of GPUs we used, doing so results in very long latencies that make it infeasible to balance latency and throughput. Moreover, running on the smaller GPU clusters requires the use of very small batch sizes, which results in poor resource utilization and low computational throughput.

\begin{table}[t]
\setlength\tabcolsep{4pt}
\centering
\small
\caption{GPU Clusters and Deployed LLMs}
\label{tab:gpu}
\begin{tabular}{@{}lccl@{}}
  \toprule
    \multirow{2}{*}{\textbf{GPU\,(Mem)}} 
      &\textbf{Cluster Size} & \textbf{Interconn.} & \multirow{2}{*}{\textbf{Model: \# GPUs}} \\
      &\scriptsize{(per node$\scriptstyle \times$\# node)} & \scriptsize{(Intra/Inter)} \\
  \midrule
  \multirow{2}{*}{A100\,(80GB)} & 16  &  \multirow{2}{*}{NVLink/Infini.} & GPT-3\,(101B): 16\\
                                & (8$\scriptstyle \times$2) &            & GPT-3\,(175B): 16\\
  \cmidrule(l{1pt}r{\lightrulewidth}){1-4}
  \multirow{5}{*}{A40\,(48GB)} &    & \multirow{5}{*}{PCIe 4.0/Infini.}& T5\,(11B):   \phantom{1234}\;8 \\
                               & 48 &                                  & OPT\,(13B): \phantom{12}\; 4\\
                               & (8$\scriptstyle \times$6) &           & GPT-3\,(39B): \phantom{\;} 16\\
                               &    &                                  & GPT-3\,(175B):\phantom{\;}32\\
                               &    &                                  & GPT-3\,(341B):\phantom{\;}48\\                              
  \bottomrule
\end{tabular}
\end{table}

\noindent
{\bf Baseline and Other Compared Systems.}
As the baseline for the evaluation, we used FasterTransformer (FT)\,\cite{nvidia2021fastertransformer}, an efficient system for LLM inference. FT supports pipeline and tensor parallel execution, as well as their combinations. For our evaluation of FT, we used the configuration that maximizes tensor parallelism for GPUs on the same machine. For example, when running with eight GPUs on one machine, we run only with tensor parallelism, and with sixteen GPUs on two machines, we run inference with two pipeline stages. This setting for FT is the same as that used in ORCA\,\cite{yu2022orca}.

We also compared the performance of ExeGPT with that of DSI\,\cite{aminabadi22dsi}, ORCA\,\cite{yu2022orca}, and vLLM\,\cite{vllm2023PagedAttention}. For these systems, we used the same parallel configuration as FT, maximizing tensor parallelism for GPUs on the same machine, which is the setting that the authors used for their evaluation. For the evaluation of ORCA we used vLLM's iteration-level scheduling mode, as ORCA is proprietary system and not publicly available. In its iteration-level schedule mode,
vLLM executes only one input encoding with other decoding computations in a batch, minimizing workload variance across batches and maximizing inference throughput. With vLLM's paging mechanism for efficient key/value cache management and early-termination of completed queries, vLLM's iteration-level schedule mode performs equivalently to ORCA's execution.

\begin{figure*}[t]
    \includegraphics[width=\textwidth]{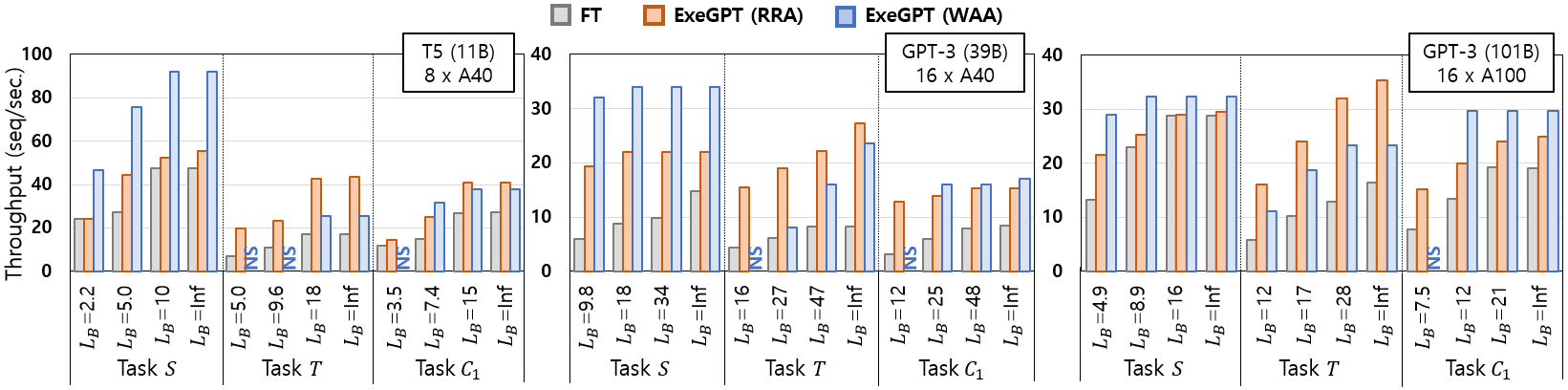}
    \caption{Throughput of ExeGPT and FT with four latency bounds ($L_B$) in seconds. $L_B$ is for generating $99^{th}$\,pctl-length sequence.}
    \label{fig:overall1}
\end{figure*}

\noindent
{\bf Evaluation Scenarios.}
We evaluated our techniques with four representative NLP tasks, namely, summarization, translation, code generation, and conversational question answering, as shown in Table\,\ref{tab:task}. Summarization is the task of generating a shorter version of the input text. Translation is the task of converting a sequence of text from one language to another. Code generation synthesizes program code from a natural language description of it. Conversational question answering requires understanding and responding to user questions in a conversational manner with various lengths of context.

To determine the sequence distribution for the tasks, we reviewed existing NLP datasets\,\cite{austin2021program, 2016WMT1, chen2021scixgen, chen2021evaluating, hendrycks2020measuring, HermannKGEKSB15, lin2021truthfulqa, reddy2019coqa, see-etal-2017-get, sharma2019bigpatent, zellers2019hellaswag, zhong2017seq2sql}. After careful examination, we found out that a truncated normal distribution (truncated below zero) provides a more accurate representation of the datasets compared to normal, log-normal, or skewed normal distributions. Therefore, we generated input/output sequences with truncated normal distribution using average and variance parameters that reflect those of the datasets corresponding to the tasks. 
In addition, we investigated the correlation of input and output lengths in these datasets. In all tasks except the translation task, the correlation between input and output sequence lengths is low, its (absolute) coefficient value ranging 0.08--0.21. For the translation task, the correlation is high (0.57--0.94), for which we can apply input length randomization across batches.

With the sequence distribution for the tasks, we generate input and output sequences for the evaluation. To enforce the sequence lengths, we made the decoding iterations continue for the given sequence lengths without emitting the end-of-sequence token, similar to the evaluation of ORCA. 
\yyy{We performed majority of the evaluation with synthesized data in this way because existing NLP datasets are not designed to evaluate LLM inference systems.
However, for a subset of our experiments, we also evaluate performance with real-world datasets and report the results for a comprehensive evaluation.}

To account for diverse service conditions and SLAs, we used four latency constraints for each task, ranging from a tight bound to a more relaxed one. To select latency constraints, we first ran the LLMs on FT with minimum to maximum batch sizes in multiples of four. We then used the bottom 10\%, 30\%, and 70\% of latencies, as well as infinity, which varied for different models and tasks. We specify these bounds for all experiments. When selecting the bounds, we used the sequence length at the $99^{th}$ percentile (pctl in short) in the distribution. Generally, this corresponds to the longest response time at the $99^{th}$ pctl. For each system, we determined scheduling parameters to ensure that the worst-case execution satisfies the latency constraint. That is, for FT and DSI, which do not apply early termination of completed queries, we applied the latency bound to generating an output sequence of maximum length. For ORCA, vLLM, and ExeGPT, we applied the same latency bound to generating the sequence length at the $99^{th}$ pctl in the distribution.

\begin{table}[t]
\setlength\tabcolsep{2pt}
\centering
\small
\caption{Evaluated NLP Tasks and Configurations}
\label{tab:task}
\begin{tabular}{@{}lccc@{}}
  \toprule
  \multirow{2}{*}{\;\textbf{Task}} & \multirow{2}{*}{\textbf{Task ID}}  
                                          & \textbf{Input Length} & \textbf{Output Length} \\
                                   &      & \scriptsize{(Avg., Std., Max)}  &  \scriptsize{(Avg., Std., $99^{th}$, Max)} \\
  \midrule
  \;Summarization & \;$S$ & (256, 252, 512) & (32, 13, 63, 80) \\
  \midrule
  \;Translation & \;$T$ & (128, 81, 256) & (128, 68, 292, 320) \\
  \midrule
  \;Code Generation& \;$G$ & (64, 23, 128) & (192, 93, 417, 480) \\
  \midrule
  \;Conversational & \;$C_1$ & (256, 115, 512) & (64, 30, 137, 160) \\
  \;\,Q\,\&\,A& \;$C_2$ & (512, 252, 1024) & (256, 134, 579, 640) \\
  \bottomrule
\end{tabular}
\end{table}

\begin{figure}[b]
\centering
    \includegraphics[width=0.47\textwidth]{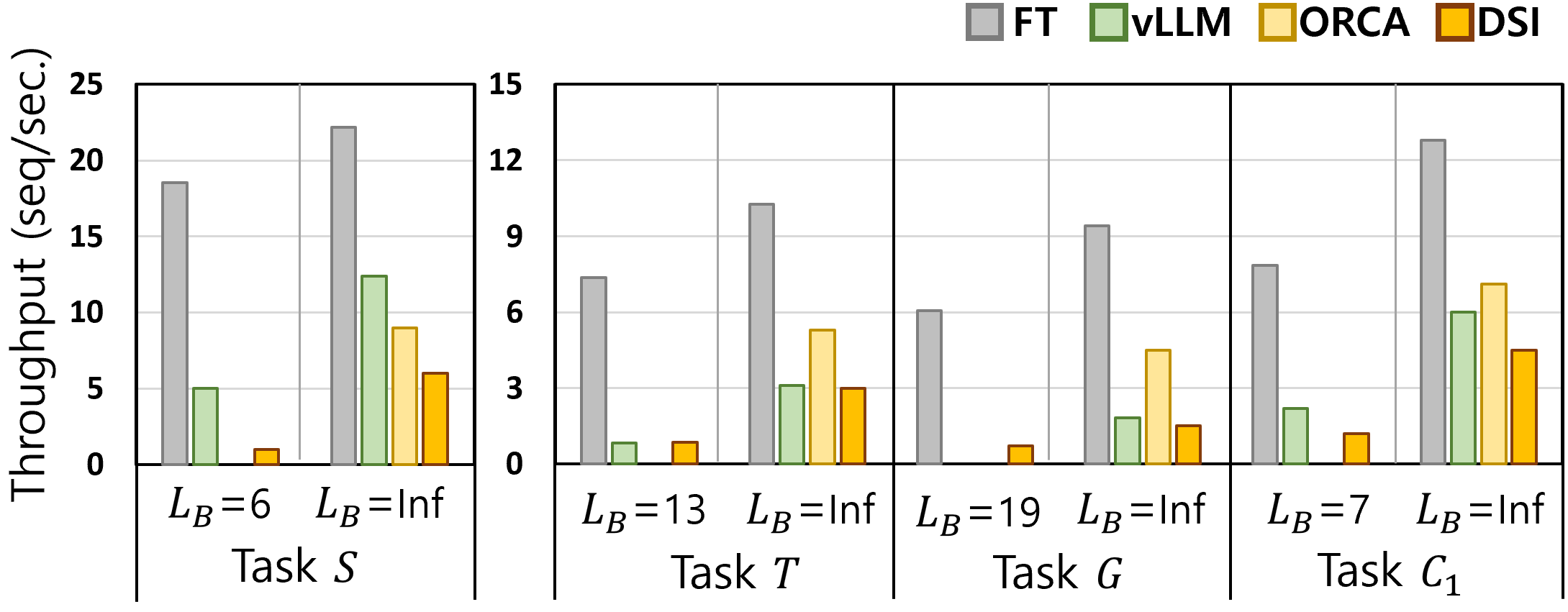}
    \caption{Throughput comparison of LLM inference systems.}
    \label{fig:opt13}
\end{figure}

\begin{figure*}[t]
    \includegraphics[width=0.99\textwidth]{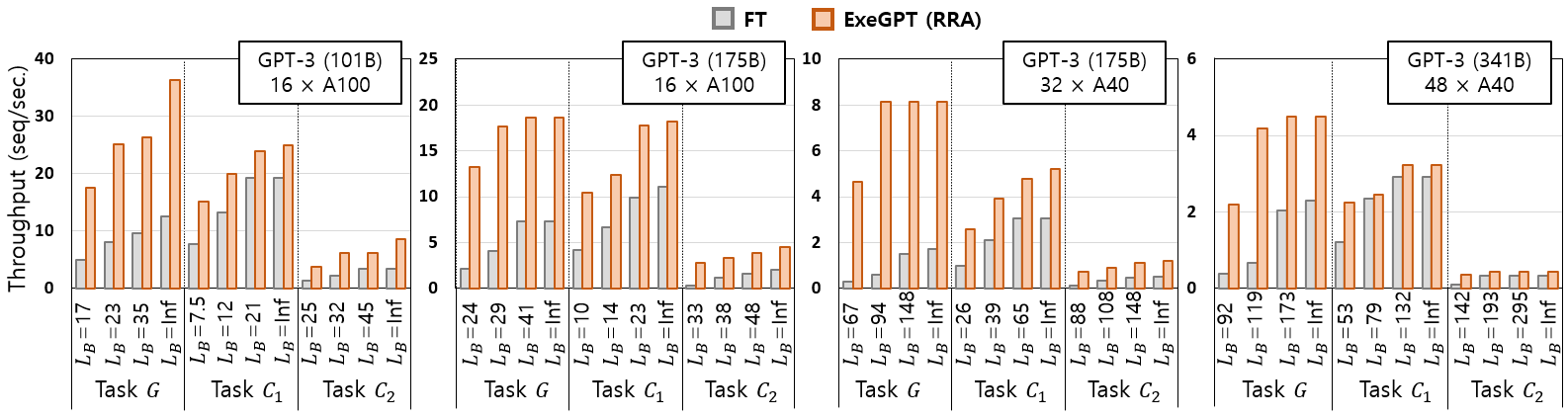}
    \caption{Throughput of ExeGPT and FT with four latency bounds ($L_B$) in seconds. $L_B$ is for generating $99^{th}$\,pctl-length sequence.}
    \label{fig:overall2}
\end{figure*}
\subsection{Performance Comparison of Existing Systems}
\label{sec:eval-compare}

We first evaluated the performance of existing systems: FT, DSI, ORCA, and vLLM. Because DSI and vLLM (their public versions) only support tensor parallelism, but not pipeline parallelism, we conducted the evaluation using a small LLM (OPT 13B) with four A40 GPUs.

Figure\,\ref{fig:opt13} shows the evaluation results. We first noticed that FT's performance is higher than those of DSI, vLLM, and ORCA for all tasks and latency bounds. While vLLM runs with larger batch sizes than FT, its executor is implemented in Python and certain execution overhead that is not masked by GPU kernels degrades its performance. Moreover, while ORCA's iteration-level scheduling improves the throughput at times, it also increases overall latency, making it hard to meet latency bounds. With FT outperforming existing systems, we subsequently compare ExeGPT to FT for further evaluation.

We conducted two separate sets of evaluations, one with small to mid-sized LLMs (11B--101B) and another with larger LLMs (101B--341B). This is because of different performance trade-off of RRA and WAA with small and large LLMs. Moreover, some tasks are known to work well with small LLMs while others require larger ones\,\cite{brown2020language, chowdhery2022palm, radford2019language, raffel2020exploring}. This section presents the evaluation results with small to mid-sized LLMs; the results with larger LLMs are shown in the next section.

\vspace*{-8pt}
\subsection{Performance Evaluation of Small to Mid-sized LLMs}
\label{sec:eval-small-llm}

With small to mid-sized LLMs, we used task $S$, $T$, and $C_1$ for the evaluation as they are well-suited for these models\,\cite{brown2020language, chowdhery2022palm, radford2019language, raffel2020exploring} (we assumed that task $C_1$ represents a relatively simple conversational task with short responses). Figure\,\ref{fig:overall1} shows the throughput of ExeGPT and FT for these tasks with four latency constraints. Overall, ExeGPT (faster of RRA and WAA) is on average 2$\times$ faster than FT across all models, tasks, and latency bounds. The highest throughput gain is 5.4$\times$ (GPT-3 39B, task $S$, $L_B$=9.8, WAA). In this specific configuration, the achieved latency is 7.8, which is a 6$\times$ improvement than FT's latency of 47 ($L_B$=Inf) while maintaining equivalent throughput. 

Between the two ExeGPT schedules, WAA outperforms RRA for tasks with relatively short output length, such as task $S$ and $C_1$. These tasks require smaller key/value cache sizes, reducing WAA's memory overhead and enhancing its efficiency. 
The addition of vLLM's paging mechanism can further enhance WAA's performance. Conversely, for task $T$ with longer output sequences, RRA generally exhibits higher throughput with better memory efficiency.
Furthermore, for a subset of the experiments WAA could not satisfy the latency bounds (denoted by NS) because it requires minimum two pipeline stages for encoding and decoding, which may lead to longer latency; also, WAA's consistent decoding batch size is $B_E\cdot S_D$ (explained in Section\,\ref{sec:decoupled-scheduling}), which increases its latency for tasks with long output sequences.

\begin{figure}[t]
\centering
    \includegraphics[width=0.48\textwidth]{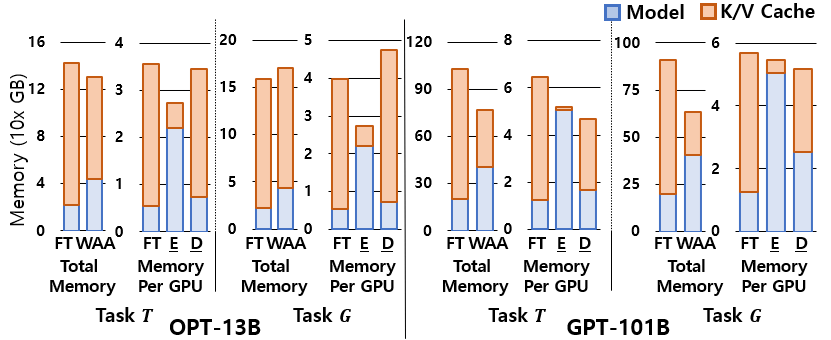}
    \caption{Memory usage of FT and WAA (\underline{E}ncoder/\underline{D}ecoder GPUs)}
    \label{fig:memory_overhead}
\end{figure}

\noindent
{\bf Memory Overhead of WAA.} We measured the memory overhead of WAA for decoder-only models and also its imbalanced memory consumption in encoder/decoder GPUs. 
Since this overhead has the most significant impact on performance when the latency constraint is at its longest (as it enables larger batch sizes and thus higher memory usage), we present the overhead results for infinite latency cases for OPT 13B and GPT-3 101B.
Figure\,\ref{fig:memory_overhead} shows the memory consumption of FT and WAA for task $T$ and $G$ (with similar results observed for other tasks).

In general, WAA consumes more model memory while using less k/v cache memory compared to FT. For OPT, WAA uses 18\% more memory for the model than FT; for GPT-3, it uses 29\% more model memory. The memory consumption between encoder and decoder GPUs is balanced for all but one as WAA-M is used in those cases. The exception is OPT with task $G$, for which WAA-C and WAA-M result in the same GPU allocation of one encoder and three decoder GPUs.

\subsection{Performance Evaluation of Large LLMs}
\label{sec:eval-large-llm}

We now explain the evaluation with larger LLMs (101B--341B). We used the code generation and conversational Q/A tasks ($G$, $C_1$, and $C_2$) as they are known to require large LLMs\,\cite{brown2020language, chowdhery2022palm, radford2019language, raffel2020exploring, touvron2023llama, touvron2023llama2}. Because of its memory overhead, WAA failed to execute for 175B and 341B models and thus we excluded WAA Schedule from the experiments. 

Figure\,\ref{fig:overall2} shows the results comparing the throughput of ExeGPT (RRA) and FT. Across all tasks and models, ExeGPT is 3.2$\times$ faster than FT on average. The performance gain of ExeGPT over FT ranges 1.1--15.2$\times$.
The gain is highest when the latency bound is tightest (GPT-3 175B and task $G$) due to our constraint-aware scheduling. However, even when the latency bound is infinity, ExeGPT achieves much higher throughput as we maintain larger decoder batches and workload. For the infinity latency bound, ExeGPT is 2.2$\times$ faster than FT on average.

\begin{figure}[t]
\centering
    \includegraphics[width=0.48\textwidth]{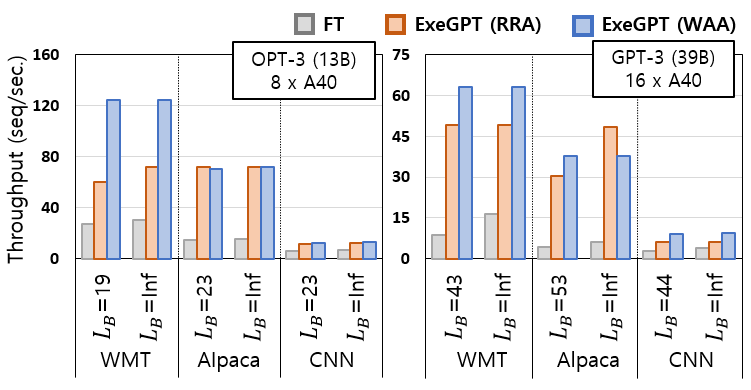}
    \caption{Throughput of ExeGPT and FT for real-world datasets (WMT, Alpaca, and CNN) with two latency bounds ($L_B$) in seconds.}
    \label{fig:real_dataset}
\end{figure}

\subsection{Evaluation with Real-World Datasets}
We evaluated the performance of FT and ExeGPT with real-world datasets using small to mid-sized LLMs (OPT 13B and GPT-3 39B). Specifically, we used the WMT dataset\,\cite{2016WMT1} for English-to-German translation, the Alpaca dataset\,\cite{taori2023stanford} for conversational Q/A, and the CNN/DailyMail dataset\,\cite{see2017get} for summarization tasks.
\textcolor{black}{We evaluated performance by enforcing the sequence lengths from these real-world datasets in a similar manner to the synthetic datasets as we explained in Section 7.1 under {\it Evaluation Scenarios}.}
We used 10\% of the datasets to estimate the sequence lengths and evaluated performance with the remaining 90\% of the datasets.

Figure\,\ref{fig:real_dataset} shows the evaluation results.
We observed that ExeGPT (the faster of RRA and WAA) performs on average 4.4$\times$ faster than FT with a maximum speedup of 8.7$\times$.
Between RRA and WAA Scheduling, WAA performs better for WMT and CNN that have relatively short output sequence lengths.
This aligns with our discussion in Section\,\ref{sec:decoupled-scheduling},
where we explained how RRA's pipeline bubble issue is exacerbated in such cases, resulting in lower throughput. 
However, for the Alpaca dataset with longer output lengths, RRA outperforms WAA, mainly due to the memory overhead associated with storing two copies of the models.
Overall, ExeGPT outperforms FT more significantly with real-world datasets than synthetic datasets (presented in Section\,\ref{sec:eval-small-llm}) due to the long-tail distribution towards long output sequences in the real-world datasets, exacerbating the diminishing decoding batch problem.

\subsection{Scheduling with Limited Information on Distribution}
\label{sec:eval-limited-seq-dist}

To find and execute with the optimal schedule, ExeGPT relies on prior knowledge of sequence length distribution, which may not always be available. 
In such cases, we evaluated ExeGPT's performance when the actual distribution differs from the one used for scheduling.
Specifically, we evaluated performance by changing three statistics in the actual output distribution: the average, standard deviation, and skewness.

For the evaluation, we used the translation task in Table\,\ref{tab:task} and OPT 13B executed on four A40 GPUs. The latency constraint was set to the bottom 30\% latency of FT. After we determined the optimal schedule for the given distribution, we changed each of the three statistics (with the other two unchanged) and then measured the throughput/latency using the selected schedule.
Note that our evaluation exclusively \textcolor{black}{focuses} on WAA because it requires 
layer re-allocation for adjusting its execution schedule to accommodate changes in distribution. In contrast, RRA can adapt simply by adjusting its encoding frequency ($N_D$) without the re-allocation.

We assumed two specific SLAs for the latency constraint: (a) ensuring that 99\% of all queries are completed within a given timeframe and (b) ensuring that a sequence of a pre-specified length (e.g. $99^{th}$\,pctl-length) is completed within a given time constraint.
To the best of our knowledge, no LLM service providers currently define explicit latency constraints in their SLAs. Therefore, for the purposes of this evaluation, we have introduced SLA-(a) and SLA-(b) as described above.

\begin{figure}[t]
\centering
    \includegraphics[width=0.475\textwidth]{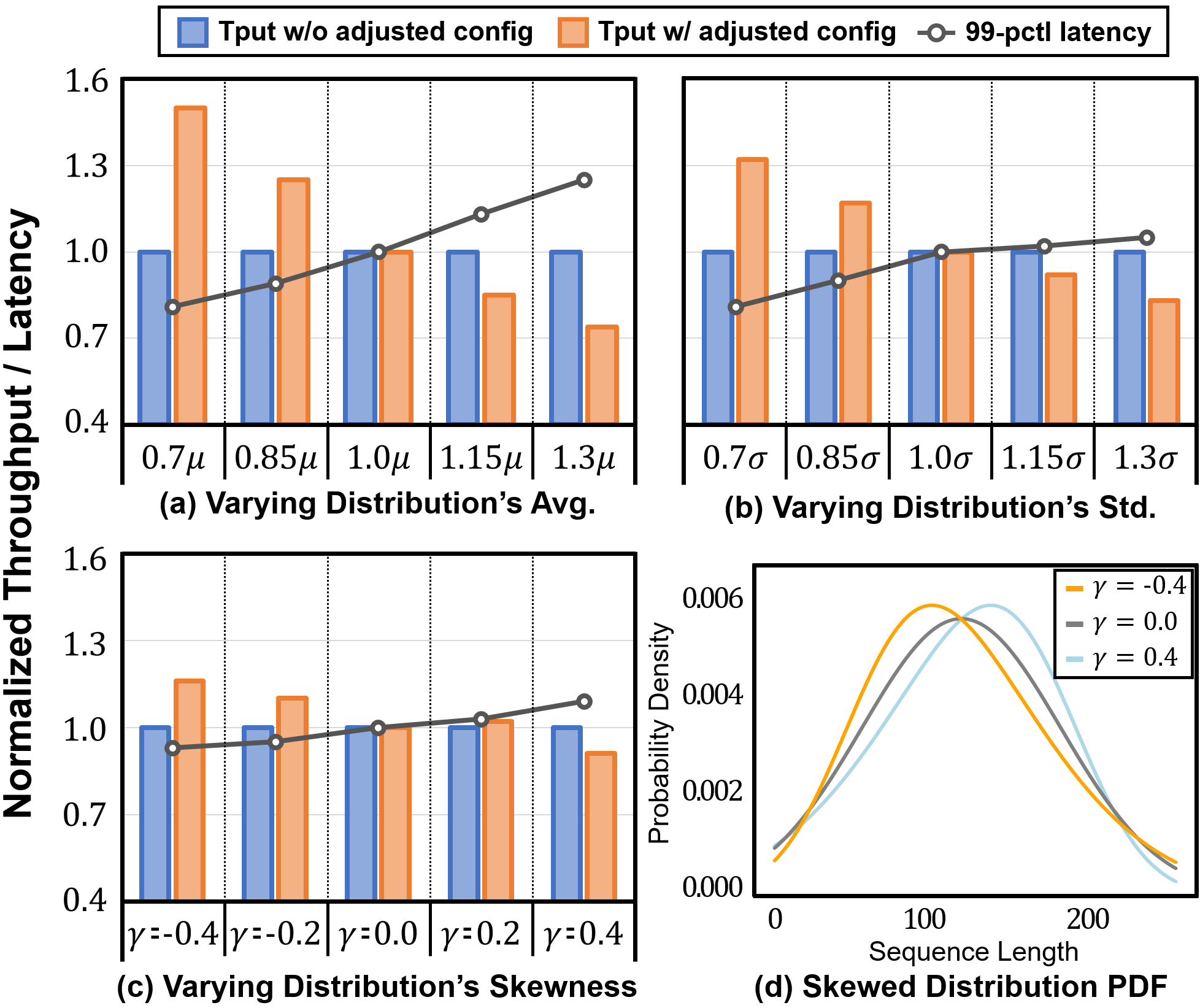}
    \caption{
        Performance of WAA with changing sequence distribution. 
        The line plot is $99^{th}$\,pctl latency, and blue bars are throughput for the execution with the same schedule. Orange bars are the throughput of the optimized execution for each changed distribution.}
    \label{fig:change_stats}
\end{figure}

Figure\,\ref{fig:change_stats}(a) shows the results for changing the average length of the distribution. The blue bars represent the throughput of the schedule for the distribution with average=$\mu$ (referred to as non-adjusted schedule), while the orange bars represent the throughput of the optimal schedule for each altered distribution. The gray line is the $99^{th}$ percentile latency of all queries for each distribution, normalized to that of the distribution with $\mu$.
We can observe that when the average length increases, the throughput of the non-adjusted schedule surpasses that of the optimal schedule.
However, this comes at the cost of not satisfying the latency constraint. Conversely, when the length decreases, the situation is reversed, with the optimal schedule outperforming the non-adjusted schedule while meeting the latency constraint with a large margin. If the latency constraint is infinity (not shown in the figure), the throughput difference between the non-adjusted and optimal schedules is considerably smaller, at most 20\% (when $\mu$=0.7).

The gray line in the figure is the 99th pctl latencies and indicates the margin that needs to be considered when scheduling with inaccurate or changing distribution information. For example, if the average length is expected to increase by 15\%, scheduling under SLA-(a) requires 13\% tighter latency constraints. Similarly, under SLA-(b), scheduling requires 10\% tighter latency constraints (not shown in the figure).

In cases where the distribution's standard deviation differs from the one used for scheduling, Figure\,\ref{fig:change_stats}(b) shows the results. Compared to Figure\,\ref{fig:change_stats}(a), the throughput and latency of the non-adjusted schedule are closer to those of the optimal ones. Especially its $99^{th}$ percentile latencies are similar to those of the optimal ones. Under SLA-(a), only a 5\% margin in latency is required for the $1.3\sigma$ case, and under SLA-(b) no margin is needed (not shown in the figure).

We now consider the skewness. 
We varied the skewness using the skew normal distribution, which generalizes the normal distribution. 
Figure\,\ref{fig:change_stats}(d) shows these skewed distribution with skewness of 0, $\pm$0.2, and $\pm$0.4\footnote{The skewness in skew normal distribution is limited to interval (-1, 1).}. The one with zero skewness is the original distribution that we use for the scheduling. Note that the average and standard deviation are equivalent among the distributions.
Figure\,\ref{fig:change_stats}(c) shows the evaluation results using these distributions. 
Compared to the other cases (changing avg. and std.)
skewness has a relatively minor impact on the throughput, with only slight differences between the non-adjusted schedule and the optimal schedules.
However, the $99^{th}$ pctl latencies are somewhat affected by skewness due to the increased long tail in the distribution.

When conducting similar experiments with RRA Scheduling, the differences in throughput and latency between the adjusted (optimal) and non-adjusted schedules are smaller compared to WAA Scheduling. For instance, when the average length changes, the throughput difference is under 28\%, and the latency difference is below 12.6\%. For changes in the other two statistics, the maximum difference is 11\% in throughput and 17\% in latency.

\subsection{Cost of Profiling, Scheduling, and (Re-)Deploying }
To run inference, we must undergo the profiling and scheduling steps, and thus, we evaluate the cost of these steps. During profiling, we measure the execution times of a single encoder/decoder with varying the control variables. This step is done once per LLM and GPU cluster, while the scheduling needs to be executed when the sequence distribution changes to ensure the latency bound is satisfied. 
For the evaluated models the profiling takes less than two hours~(on {\footnotesize AMD~EPYC~7313}).
Scheduling for the models takes three seconds to two minutes for RRA and one to five minutes for WAA. In comparison, scheduling with exhaustive search needs a minimum of five hours and can extend to an entire day.

Deploying a model instance according to the selected schedule involves loading the model from SSD to GPU memory in parallel. 
In the case of distribution changes, re-deploying the model, to adjust resource allocation accordingly, requires reloading the model either from CPU DRAM or SSD. We measured the costs of deploying and re-deploying and summarized the results in Table\,\ref{tab:deploy_cost}.
The deployment cost, i.e., loading the model from SSD, ranges from 2.1 to 15.1 seconds across small to large LLMs. Re-deploying the model to load the model from CPU's DRAM takes 0.9 to 3.5 seconds.
We consider these re-deployment costs to be reasonably modest in real-world LLM serving scenarios.

\begin{table}[t]
\setlength\tabcolsep{4pt}
\centering
\small
\caption{The cost of loading LLMs from SSD or CPU's DRAM}
\label{tab:deploy_cost}
\setlength{\fboxrule}{1.5pt}

\begin{tabular}{@{}ccrr@{}}
  \toprule
  \multirow{1}{*}{\textbf{Model}} & \multirow{1}{*}{\textbf{\#GPUs}} & \textbf{Loading from DRAM} & \textbf{Loading from SSD} \\
  \midrule
  39B & 16 & 0.9 secs.\phantom{0123} & 2.1 secs. \phantom{0123}\\
  \midrule
  101B & 32 & 1.3 secs.\phantom{0123} & 7.1 secs. \phantom{0123}\\
  \midrule
  175B & 32 & 2.1 secs.\phantom{0123} & 11.9 secs. \phantom{0123}\\
  \midrule
  341B & 48 & 3.5 secs.\phantom{0123} & 15.1 secs. \phantom{0123}\\
  \bottomrule
\end{tabular}

\end{table}

\begin{table}[tbh!]
\setlength\tabcolsep{3.5pt}
\centering
\small
\renewcommand{\arraystretch}{1.2}
\caption{Percentage of non-monotonic points$^*$}
\label{tab:monotonicity}
\begin{tabular}{@{}ccccccc@{}}

\toprule
\multirow{2}{*}{\textbf{\!Task\!}} & \multirow{2}{*}{\textbf{Tol.$^\dagger$}} & \multicolumn{2}{c}{\textbf{RRA}}                       & \multicolumn{3}{c}{\textbf{WAA}} \\
                                   &                                & \multicolumn{1}{c}{$B_E$} & \multicolumn{1}{c}{$N_D$} & \multicolumn{1}{c}{$B_E$} & \multicolumn{1}{c}{TP} & \multicolumn{1}{c}{$B_m$} \\

\hline
\multirow{3}{*}{S}    & 2\%  & (0.6, 2.3) & (0.0, 0.0) & (0.0, 0.0) & (0.0, 0.0) & (0.8, 0.0) \\
                      & 5\%  & (0.6, 0.3) & (0.0, 0.0) & (0.0, 0.0) & (0.0, 0.0) & (0.5, 0.0) \\
                      & 10\% & (0.3, 0.0) & (0.0, 0.0) & (0.0, 0.0) & (0.0, 0.0) & (0.4, 0.0) \\

\hline
\multirow{3}{*}{T}    & 2\% & (0.2, 0.1) & (0.0, 0.0) & (0.1, 0.0) & (1.2, 0.0) & (10.9, 0.0) \\
                      & 5\% & (0.2, 0.0) & (0.0, 0.0) & (0.0, 0.0) & (0.8, 0.0) & (10.6, 0.0) \\
                      & 10\% & (0.2, 0.0) & (0.0, 0.0) & (0.0, 0.0) & (0.4, 0.0) & (10.4, 0.0) \\

\bottomrule
\multicolumn{7}{r}{\scriptsize $*$ Each cell represents (Latency, Throughput).} \\
\multicolumn{7}{r}{\scriptsize $\dagger$ Tolerance values are calculated as \% of $L_B$ ($70^{th}$ pctl) and the achieved tput.} \\
\end{tabular}
\end{table}

\subsection{Monotonicity and Throughput/Latency Trade-off}

To assess the monotonicity, we run experiments with T5 and GPT-3 (39B and 101B) and with all tasks. In the experiments, we adjusted the variables individually, observing the corresponding changes in throughput and latency. That is, we swept through a range of values for a variable while keeping the others fixed, and repeated the measurement for all combinations of those other variables. Then we quantify the fraction of the points in the range that the monotonicity property holds.

Table\,\ref{tab:monotonicity} shows the results for GPT-3 39B. We noted that the control variables generally exhibit monotonic behavior with throughput and latency. 
With 5\% tolerance values, 97\% of points in total (including all control variables in RRA and WAA) show monotonicity in throughput, and 96\% in latency.

Decoder micro-batch ($B_m$) has slightly more non-monotonic points for task $T$ due to its potential interference with partial tensor-parallelism. The two control variables, when used in combination, may significantly reduce the workload to a very small size, resulting in non-monotonic execution latencies. After excluding those points, less than 5\% of its remaining points show a non-monotonic behavior.

\begin{table}[t]
\centering
\small
\caption{Selected Schedule and the Optimal Configurations}
\label{tab:casestudy}
\begin{tabular}{@{}ccccc@{}}
  \toprule
  \multirow{2}{*}{\textbf{$L_B$}} & \textbf{Selected} & \multirow{2}{*}{\textbf{Control Variables}} & \textbf{Latency} & \textbf{Tput} \\
 & \textbf{Schedule} &  & \textbf{(sec.)} & \textbf{(seq./sec.)} \\
  \midrule
  3.1 & WAA & $B_E$=4,  $T_P$=2 & 3.01 & 22.41 \\
  \midrule
  5.9 & WAA & $B_E$=10, $T_P$=2 & 4.11 & 23.76 \\
  \midrule
  11.5 & RRA & $B_E$=52, $N_D$=13, $T_P$=4 & 10.23 & 25.23 \\
  \midrule
  $Inf.$ & RRA & $B_E$=49, $N_D$=7, $T_P$=4 & 16.87 & 26.15 \\
  \bottomrule
\end{tabular}
\end{table}

\noindent
{\bf Case Study.} We looked into throughput/latency trade-off of our scheduling for different latency bounds with OPT 13B and task $S$. Table \ref{tab:casestudy} summarizes the selected values of the control variables and schedules with four different latency bounds. 
As the bound $L_B$ relaxes, different control variables are adjusted: first, encoder batch size $B_E$ increases, then the schedule shifts from WAA to RRA. When $L_B$ approaches infinity, encoding frequency doubles (or $N_D$ halves). The shortest $L_B$ yields 80\% of the maximum throughput demonstrating efficient trade-off by our scheduler.

\subsection{Variance of Encoder and Decoder Workload}

\label{sec:eval-workload-var}
To ensure the selected schedule meets the latency constraint, it is essential to maintain encoder/decoder workload consistently within a fixed boundary. For this purpose, we dynamically adjust the workload at runtime, but even without this adjustment the variance of the workload is evaluated to be relatively small. To analyze the workload variance, we examined the execution of the selected schedules for the evaluated models and tasks.

Table\,\ref{tab:var_workload} summarizes the results for OPT and task $S$ that are representative of others not shown due to the space limit. It shows the $99^{th}$ percentile ranges of encoder/decoder's single stage execution times.
While the encoder workload variance is not negligible, its impact on latency remains minor due to the asynchronous execution of encoder and decoder stages in WAA and input workload randomization across batches in RRA. We do consider this variance when determining control variables to satisfy latency constraints. On the other hand, the variance of decoder execution time is notably low, less than 6\%. The combination of this small workload variance and our dynamic workload adjustment ensures a consistent encoder/decoder workload throughout executions.

\section{Conclusion}

This paper introduces ExeGPT, a system for executing LLM inference that maximizes throughput under a given latency bound. The system utilizes two scheduling strategies, based on RRA and WAA policies, that have four control variables, enabling flexible trade-offs between throughput and latency. To efficiently find the optimal values of the variables, we implemented a scheduling algorithm based on the branch-and-bound method, exploiting their monotonicity properties.

We evaluated ExeGPT on six LLM instances of T5, OPT, and GPT-3 and five NLP tasks, each with four distinct latency constraints. The evaluation shows that ExeGPT achieves up to 15.2$\times$ higher throughput when latency is the same and 6$\times$ better latency when throughput is the same than the state-of-the-art FasterTransformer. On average, ExeGPT yields a throughput gain of 2.9$\times$ across the twenty evaluation scenarios. We also found that when adapting to changing sequence distributions, the cost of adjusting the schedule in ExeGPT is reasonably modest.
In summary, ExeGPT provides an effective solution for optimizing and executing LLM inference for diverse NLP workload and serving conditions.

\begin{table}[t]
\centering
\small
\caption{Variance of encoder/decoder exec. times in seconds}
\label{tab:var_workload}
\begin{tabular}{@{}ccc@{}}
  \toprule
 \textbf{Schedule} & \textbf{Encoder} \footnotesize{($99^{th}$\,pctl Range)} & \textbf{Decoder} \footnotesize{($99^{th}$\,pctl Range)}\\

  \midrule
RRA & 1.68 ($\pm$0.12, $\pm$7.1\%) &  0.038 ($\pm$0.001, $\pm$2.4\%) \\

WAA & 0.85 ($\pm$0.10, $\pm$11.8\%)  & 0.041 ($\pm$0.002, $\pm$5.5\%) \\

  \bottomrule
\end{tabular}
\end{table}

\section*{Acknowledgement}
This work is supported by 
Institute of Information \& communications Technology Planning \& Evaluation (IITP) grant funded by the Korea government (MSIT)
(No.2020-0-01373, No.2021-0-00310, 
No.2022-0-00498, and No.2024-2021-0-01817).
Jiwon Seo is the corresponding author.

\bibliographystyle{plain}
\bibliography{paper}

\begin{thebibliography}{10}

\bibitem{aminabadi22dsi}
Reza~Yazdani Aminabadi, Samyam Rajbhandari, Ammar~Ahmad Awan, Cheng Li, Du~Li, Elton Zheng, Olatunji Ruwase, Shaden Smith, Minjia Zhang, Jeff Rasley, and Yuxiong He.
\newblock Deepspeed-inference: Enabling efficient inference of transformer models at unprecedented scale.
\newblock In {\em Proceedings of the International Conference on High Performance Computing, Networking, Storage and Analysis}, SC '22. IEEE Press, 2022.

\bibitem{austin2021program}
Jacob Austin, Augustus Odena, Maxwell Nye, Maarten Bosma, Henryk Michalewski, David Dohan, Ellen Jiang, Carrie Cai, Michael Terry, Quoc Le, et~al.
\newblock Program synthesis with large language models.
\newblock {\em arXiv preprint arXiv:2108.07732}, 2021.

\bibitem{2016WMT1}
Ond~{r}ej Bojar, Rajen Chatterjee, Christian Federmann, Yvette Graham, Barry Haddow, Matthias Huck, Antonio Jimeno~Yepes, Philipp Koehn, Varvara Logacheva, Christof Monz, Matteo Negri, Aurelie Neveol, Mariana Neves, Martin Popel, Matt Post, Raphael Rubino, Carolina Scarton, Lucia Specia, Marco Turchi, Karin Verspoor, and Marcos Zampieri.
\newblock Findings of the 2016 conference on machine translation.
\newblock In {\em Proceedings of the First Conference on Machine Translation}, pages 131--198, Berlin, Germany, August 2016. Association for Computational Linguistics.

\bibitem{bondarenko2021understanding}
Yelysei Bondarenko, Markus Nagel, and Tijmen Blankevoort.
\newblock Understanding and overcoming the challenges of efficient transformer quantization.
\newblock {\em arXiv preprint arXiv:2109.12948}, 2021.

\bibitem{brown2020language}
Tom Brown, Benjamin Mann, Nick Ryder, Melanie Subbiah, Jared~D Kaplan, Prafulla Dhariwal, Arvind Neelakantan, Pranav Shyam, Girish Sastry, Amanda Askell, et~al.
\newblock Language models are few-shot learners.
\newblock {\em Advances in neural information processing systems}, 33:1877--1901, 2020.

\bibitem{chen2021scixgen}
Hong Chen, Hiroya Takamura, and Hideki Nakayama.
\newblock Scixgen: A scientific paper dataset for context-aware text generation.
\newblock {\em arXiv preprint arXiv:2110.10774}, 2021.

\bibitem{chen2021evaluating}
Mark Chen, Jerry Tworek, Heewoo Jun, Qiming Yuan, Henrique Ponde de~Oliveira Pinto, Jared Kaplan, Harri Edwards, Yuri Burda, Nicholas Joseph, Greg Brockman, et~al.
\newblock Evaluating large language models trained on code.
\newblock {\em arXiv preprint arXiv:2107.03374}, 2021.

\bibitem{chowdhery2022palm}
Aakanksha Chowdhery, Sharan Narang, Jacob Devlin, Maarten Bosma, Gaurav Mishra, Adam Roberts, Paul Barham, Hyung~Won Chung, Charles Sutton, Sebastian Gehrmann, et~al.
\newblock Palm: Scaling language modeling with pathways.
\newblock {\em arXiv preprint arXiv:2204.02311}, 2022.

\bibitem{dao2022flashattention}
Tri Dao, Dan Fu, Stefano Ermon, Atri Rudra, and Christopher R{\'e}.
\newblock Flashattention: Fast and memory-efficient exact attention with io-awareness.
\newblock {\em Advances in Neural Information Processing Systems}, 35:16344--16359, 2022.

\bibitem{dettmers2022llm}
Tim Dettmers, Mike Lewis, Younes Belkada, and Luke Zettlemoyer.
\newblock Llm. int8 (): 8-bit matrix multiplication for transformers at scale.
\newblock {\em arXiv preprint arXiv:2208.07339}, 2022.

\bibitem{frantar2022gptq}
Elias Frantar, Saleh Ashkboos, Torsten Hoefler, and Dan Alistarh.
\newblock Gptq: Accurate post-training quantization for generative pre-trained transformers.
\newblock {\em arXiv preprint arXiv:2210.17323}, 2022.

\bibitem{hendrycks2020measuring}
Dan Hendrycks, Collin Burns, Steven Basart, Andy Zou, Mantas Mazeika, Dawn Song, and Jacob Steinhardt.
\newblock Measuring massive multitask language understanding.
\newblock {\em arXiv preprint arXiv:2009.03300}, 2020.

\bibitem{HermannKGEKSB15}
Karl~Moritz Hermann, Tomás Kociský, Edward Grefenstette, Lasse Espeholt, Will Kay, Mustafa Suleyman, and Phil Blunsom.
\newblock Teaching machines to read and comprehend.
\newblock In {\em NIPS}, pages 1693--1701, 2015.

\bibitem{huang2019gpipe}
Yanping Huang, Youlong Cheng, Ankur Bapna, Orhan Firat, Dehao Chen, Mia Chen, HyoukJoong Lee, Jiquan Ngiam, Quoc~V Le, Yonghui Wu, et~al.
\newblock Gpipe: Efficient training of giant neural networks using pipeline parallelism.
\newblock {\em Advances in neural information processing systems}, 32, 2019.

\bibitem{vllm2023PagedAttention}
Woosuk Kwon, Zhuohan Li, Siyuan Zhuang, Ying Sheng, Lianmin Zheng, Cody Yu, Joey Gonzalez, Hao Zhang, and Ion Stoica.
\newblock vllm: Easy, fast, and cheap llm serving with pagedattention, 2023.

\bibitem{li21chimera}
Shigang Li and Torsten Hoefler.
\newblock Chimera: Efficiently training large-scale neural networks with bidirectional pipelines.
\newblock In {\em Proceedings of the International Conference for High Performance Computing, Networking, Storage and Analysis}, SC '21, New York, NY, USA, 2021. Association for Computing Machinery.

\bibitem{li2021terapipe}
Zhuohan Li, Siyuan Zhuang, Shiyuan Guo, Danyang Zhuo, Hao Zhang, Dawn Song, and Ion Stoica.
\newblock Terapipe: Token-level pipeline parallelism for training large-scale language models.
\newblock In {\em International Conference on Machine Learning}, pages 6543--6552. PMLR, 2021.

\bibitem{lin2021truthfulqa}
Stephanie Lin, Jacob Hilton, and Owain Evans.
\newblock Truthfulqa: Measuring how models mimic human falsehoods.
\newblock {\em arXiv preprint arXiv:2109.07958}, 2021.

\bibitem{mishra2021accelerating}
Asit Mishra, Jorge~Albericio Latorre, Jeff Pool, Darko Stosic, Dusan Stosic, Ganesh Venkatesh, Chong Yu, and Paulius Micikevicius.
\newblock Accelerating sparse deep neural networks.
\newblock {\em arXiv preprint arXiv:2104.08378}, 2021.

\bibitem{movckus1975bayesian}
Jonas Mo{\v{c}}kus.
\newblock On bayesian methods for seeking the extremum.
\newblock In {\em Optimization Techniques IFIP Technical Conference: Novosibirsk, July 1--7, 1974}, pages 400--404. Springer, 1975.

\bibitem{narayanan19pipedream}
Deepak Narayanan, Aaron Harlap, Amar Phanishayee, Vivek Seshadri, Nikhil~R. Devanur, Gregory~R. Ganger, Phillip~B. Gibbons, and Matei Zaharia.
\newblock Pipedream: Generalized pipeline parallelism for dnn training.
\newblock In {\em Proceedings of the 27th ACM Symposium on Operating Systems Principles}, SOSP '19, page 1–15, New York, NY, USA, 2019. Association for Computing Machinery.

\bibitem{narayanan2021efficient}
Deepak Narayanan, Mohammad Shoeybi, Jared Casper, Patrick LeGresley, Mostofa Patwary, Vijay Korthikanti, Dmitri Vainbrand, Prethvi Kashinkunti, Julie Bernauer, Bryan Catanzaro, et~al.
\newblock Efficient large-scale language model training on gpu clusters using megatron-{LM}.
\newblock In {\em Proceedings of the International Conference for High Performance Computing, Networking, Storage and Analysis}, pages 1--15, 2021.

\bibitem{nvidia2021fastertransformer}
NVIDIA.
\newblock Fastertransformer, 2021.

\bibitem{ott2019fairseq}
Myle Ott, Sergey Edunov, Alexei Baevski, Angela Fan, Sam Gross, Nathan Ng, David Grangier, and Michael Auli.
\newblock fairseq: A fast, extensible toolkit for sequence modeling.
\newblock {\em arXiv preprint arXiv:1904.01038}, 2019.

\bibitem{radford2018improving}
Alec Radford, Karthik Narasimhan, Tim Salimans, Ilya Sutskever, et~al.
\newblock Improving language understanding by generative pre-training.
\newblock 2018.

\bibitem{radford2019language}
Alec Radford, Jeffrey Wu, Rewon Child, David Luan, Dario Amodei, Ilya Sutskever, et~al.
\newblock Language models are unsupervised multitask learners.
\newblock {\em OpenAI blog}, 1(8):9, 2019.

\bibitem{rae2021scaling}
Jack~W Rae, Sebastian Borgeaud, Trevor Cai, Katie Millican, Jordan Hoffmann, Francis Song, John Aslanides, Sarah Henderson, Roman Ring, Susannah Young, et~al.
\newblock Scaling language models: Methods, analysis \& insights from training gopher.
\newblock {\em arXiv preprint arXiv:2112.11446}, 2021.

\bibitem{raffel2020exploring}
Colin Raffel, Noam Shazeer, Adam Roberts, Katherine Lee, Sharan Narang, Michael Matena, Yanqi Zhou, Wei Li, and Peter~J Liu.
\newblock Exploring the limits of transfer learning with a unified text-to-text transformer.
\newblock {\em The Journal of Machine Learning Research}, 21(1):5485--5551, 2020.

\bibitem{reddy2019coqa}
Siva Reddy, Danqi Chen, and Christopher~D Manning.
\newblock Coqa: A conversational question answering challenge.
\newblock {\em Transactions of the Association for Computational Linguistics}, 7:249--266, 2019.

\bibitem{see-etal-2017-get}
Abigail See, Peter~J. Liu, and Christopher~D. Manning.
\newblock Get to the point: Summarization with pointer-generator networks.
\newblock In {\em Proceedings of the 55th Annual Meeting of the Association for Computational Linguistics (Volume 1: Long Papers)}, pages 1073--1083, Vancouver, Canada, July 2017. Association for Computational Linguistics.

\bibitem{see2017get}
Abigail See, Peter~J Liu, and Christopher~D Manning.
\newblock Get to the point: Summarization with pointer-generator networks.
\newblock {\em arXiv preprint arXiv:1704.04368}, 2017.

\bibitem{sharma2019bigpatent}
Eva Sharma, Chen Li, and Lu~Wang.
\newblock Bigpatent: A large-scale dataset for abstractive and coherent summarization.
\newblock {\em arXiv preprint arXiv:1906.03741}, 2019.

\bibitem{shazeer2020glu}
Noam Shazeer.
\newblock Glu variants improve transformer.
\newblock {\em arXiv preprint arXiv:2002.05202}, 2020.

\bibitem{shoeybi2019megatron}
Mohammad Shoeybi, Mostofa Patwary, Raul Puri, Patrick LeGresley, Jared Casper, and Bryan Catanzaro.
\newblock Megatron-lm: Training multi-billion parameter language models using model parallelism.
\newblock {\em arXiv preprint arXiv:1909.08053}, 2019.

\bibitem{tao2022compression}
Chaofan Tao, Lu~Hou, Wei Zhang, Lifeng Shang, Xin Jiang, Qun Liu, Ping Luo, and Ngai Wong.
\newblock Compression of generative pre-trained language models via quantization.
\newblock {\em arXiv preprint arXiv:2203.10705}, 2022.

\bibitem{taori2023alpaca}
Rohan Taori, Ishaan Gulrajani, Tianyi Zhang, Yann Dubois, Xuechen Li, Carlos Guestrin, Percy Liang, and Tatsunori~B Hashimoto.
\newblock Alpaca: A strong, replicable instruction-following model.
\newblock {\em Stanford Center for Research on Foundation Models. https://crfm. stanford. edu/2023/03/13/alpaca. html}, 3(6):7, 2023.

\bibitem{taori2023stanford}
Rohan Taori, Ishaan Gulrajani, Tianyi Zhang, Yann Dubois, Xuechen Li, Carlos Guestrin, Percy Liang, and Tatsunori~B Hashimoto.
\newblock Stanford alpaca: An instruction-following llama model, 2023.

\bibitem{tay2022unifying}
Yi~Tay, Mostafa Dehghani, Vinh~Q Tran, Xavier Garcia, Dara Bahri, Tal Schuster, Huaixiu~Steven Zheng, Neil Houlsby, and Donald Metzler.
\newblock Unifying language learning paradigms.
\newblock {\em arXiv preprint arXiv:2205.05131}, 2022.

\bibitem{thoppilan2022lamda}
Romal Thoppilan, Daniel De~Freitas, Jamie Hall, Noam Shazeer, Apoorv Kulshreshtha, Heng-Tze Cheng, Alicia Jin, Taylor Bos, Leslie Baker, Yu~Du, et~al.
\newblock Lamda: Language models for dialog applications.
\newblock {\em arXiv preprint arXiv:2201.08239}, 2022.

\bibitem{touvron2023llama}
Hugo Touvron, Thibaut Lavril, Gautier Izacard, Xavier Martinet, Marie-Anne Lachaux, Timoth{\'e}e Lacroix, Baptiste Rozi{\`e}re, Naman Goyal, Eric Hambro, Faisal Azhar, et~al.
\newblock Llama: Open and efficient foundation language models.
\newblock {\em arXiv preprint arXiv:2302.13971}, 2023.

\bibitem{touvron2023llama2}
Hugo Touvron, Louis Martin, Kevin Stone, Peter Albert, Amjad Almahairi, Yasmine Babaei, Nikolay Bashlykov, Soumya Batra, Prajjwal Bhargava, Shruti Bhosale, et~al.
\newblock Llama 2: Open foundation and fine-tuned chat models.
\newblock {\em arXiv preprint arXiv:2307.09288}, 2023.

\bibitem{tuy2000monotonic}
Hoang Tuy.
\newblock Monotonic optimization: Problems and solution approaches.
\newblock {\em SIAM Journal on Optimization}, 11(2):464--494, 2000.

\bibitem{tuy2006discrete}
Hoang Tuy, Michel Minoux, and NT~Hoai-Phuong.
\newblock Discrete monotonic optimization with application to a discrete location problem.
\newblock {\em SIAM Journal on Optimization}, 17(1):78--97, 2006.

\bibitem{vaswani2017attention}
Ashish Vaswani, Noam Shazeer, Niki Parmar, Jakob Uszkoreit, Llion Jones, Aidan~N Gomez, {\L}ukasz Kaiser, and Illia Polosukhin.
\newblock Attention is all you need.
\newblock {\em Advances in neural information processing systems}, 30, 2017.

\bibitem{xiao2022smoothquant}
Guangxuan Xiao, Ji~Lin, Mickael Seznec, Julien Demouth, and Song Han.
\newblock Smoothquant: Accurate and efficient post-training quantization for large language models.
\newblock {\em arXiv preprint arXiv:2211.10438}, 2022.

\bibitem{yao2022zeroquant}
Zhewei Yao, Reza Yazdani~Aminabadi, Minjia Zhang, Xiaoxia Wu, Conglong Li, and Yuxiong He.
\newblock Zeroquant: Efficient and affordable post-training quantization for large-scale transformers.
\newblock {\em Advances in Neural Information Processing Systems}, 35:27168--27183, 2022.

\bibitem{yu2022orca}
Gyeong-In Yu, Joo~Seong Jeong, Geon-Woo Kim, Soojeong Kim, and Byung-Gon Chun.
\newblock {ORCA}: A distributed serving system for transformer-based generative models.
\newblock In {\em 16th USENIX Symposium on Operating Systems Design and Implementation (OSDI 22)}, pages 521--538, 2022.

\bibitem{zafrir2019q8bert}
Ofir Zafrir, Guy Boudoukh, Peter Izsak, and Moshe Wasserblat.
\newblock Q8bert: Quantized 8bit bert.
\newblock In {\em 2019 Fifth Workshop on Energy Efficient Machine Learning and Cognitive Computing-NeurIPS Edition (EMC2-NIPS)}, pages 36--39. IEEE, 2019.

\bibitem{zafrir2021prune}
Ofir Zafrir, Ariel Larey, Guy Boudoukh, Haihao Shen, and Moshe Wasserblat.
\newblock Prune once for all: Sparse pre-trained language models.
\newblock {\em arXiv preprint arXiv:2111.05754}, 2021.

\bibitem{zellers2019hellaswag}
Rowan Zellers, Ari Holtzman, Yonatan Bisk, Ali Farhadi, and Yejin Choi.
\newblock Hellaswag: Can a machine really finish your sentence?
\newblock {\em arXiv preprint arXiv:1905.07830}, 2019.

\bibitem{zhang2022platon}
Qingru Zhang, Simiao Zuo, Chen Liang, Alexander Bukharin, Pengcheng He, Weizhu Chen, and Tuo Zhao.
\newblock Platon: Pruning large transformer models with upper confidence bound of weight importance.
\newblock In {\em International Conference on Machine Learning}, pages 26809--26823. PMLR, 2022.

\bibitem{zhang2022opt}
Susan Zhang, Stephen Roller, Naman Goyal, Mikel Artetxe, Moya Chen, Shuohui Chen, Christopher Dewan, Mona Diab, Xian Li, Xi~Victoria Lin, et~al.
\newblock Opt: Open pre-trained transformer language models.
\newblock {\em arXiv preprint arXiv:2205.01068}, 2022.

\bibitem{zhong2017seq2sql}
Victor Zhong, Caiming Xiong, and Richard Socher.
\newblock Seq2sql: Generating structured queries from natural language using reinforcement learning.
\newblock {\em arXiv preprint arXiv:1709.00103}, 2017.

\end{thebibliography}

\end{document}